\documentclass[acmsmall]{acmart}
\usepackage{CJKutf8}
\usepackage[utf8]{inputenc}
\usepackage[T1]{fontenc}
\usepackage{textcomp}
\usepackage{xcolor,colortbl}
\usepackage{multirow}
\usepackage{multicol}
\usepackage{threeparttable}
\usepackage{soul}

\AtBeginDocument{%
  \providecommand\BibTeX{{%
    \normalfont B\kern-0.5em{\scshape i\kern-0.25em b}\kern-0.8em\TeX}}}

\begin{document}

\setcopyright{acmcopyright}
\acmJournal{PACMHCI}
\acmYear{2024} \acmVolume{8} \acmNumber{CSCW} \acmArticle{01} \acmMonth{01} \acmPrice{15.00}\acmDOI{XX.XXXX/XXXXXXX}

\title[Engage Wider Audience or Facilitate Quality Answers]{Engage Wider Audience or Facilitate Quality Answers? a Mixed-methods Analysis of Questioning Strategies for Research Sensemaking on a Community Q\&A Site}

\author{Changyang He}
\affiliation{%
  \institution{Hong Kong University of Science and Technology}
  \city{Hong Kong SAR}
  \country{China}
}
\email{cheai@cse.ust.hk}

\author{Yue Deng}
\affiliation{%
  \institution{Hong Kong University of Science and Technology}
  \city{Hong Kong SAR}
  \country{China}
}
\email{ydengbi@cse.ust.hk}

\author{Lu He}
\affiliation{%
  \institution{University of Wisconsin-Milwaukee}
  \city{Milwaukee}
  \country{USA}
}
\email{he32@uwm.edu}

\author{Qingyu Guo}
\affiliation{%
  \institution{Hong Kong University of Science and Technology}
  \city{Hong Kong SAR}
  \country{China}
}
\email{qguoag@connect.ust.hk}

\author{Yu Zhang}
\affiliation{%
  \institution{City University of Hong Kong}
  \city{Hong Kong}
  \country{China}
}
\email{yui.zhang@my.cityu.edu.hk}

\author{Zhicong Lu}
\affiliation{%
  \institution{City University of Hong Kong}
  \city{Hong Kong}
  \country{China}
}
\email{zhicong.lu@cityu.edu.hk}

\author{Bo Li}
\affiliation{%
  \institution{Hong Kong University of Science and Technology}
  \city{Hong Kong SAR}
  \country{China}
}
\email{bli@cse.ust.hk}

\renewcommand{\shortauthors}{Changyang He et al.}
	
\newcommand{\gr}{\leavevmode\color{lightgray}}
\newcommand{\cgr}{\cellcolor{green!25}}
\newcommand{\crd}{\cellcolor{pink!25}}

\begin{abstract}

Discussing research-sensemaking questions on Community Question and Answering (CQA) platforms has been an increasingly common practice for the public to participate in science communication. Nonetheless, how users strategically craft research-sensemaking questions to engage public participation and facilitate knowledge construction is a significant yet less understood problem. To fill this gap, we collected 837 science-related questions and 157,684 answers from Zhihu, and conducted a mixed-methods study to explore user-developed strategies in proposing research-sensemaking questions, and their potential effects on public engagement and knowledge construction. Through open coding, we captured a comprehensive taxonomy of question-crafting strategies, such as eyecatching narratives with counter-intuitive claims and rigorous descriptions with data use. Regression analysis indicated that these strategies correlated with user engagement and answer construction in different ways (e.g., emotional questions attracted more views and answers), yet there existed a general divergence between wide participation and quality knowledge establishment, when most questioning strategies could not ensure both. Based on log analysis, we further found that collaborative editing afforded unique values in refining research-sensemaking questions regarding accuracy, rigor, comprehensiveness and attractiveness. We propose design implications to facilitate accessible, accurate and engaging science communication on CQA platforms.

\end{abstract}



\maketitle

\section{INTRODUCTION}

Science communication is the appropriate use of communication techniques to produce personal awareness, enjoyment, interest, opinion-forming, or understanding of science~\cite{burns2003science}. It provides a valuable means for the public to make sense of research progress and improve decision-making in everyday life~\cite{treise2002advancing}. Recently, the development of social media accelerated scientific knowledge dissemination and exchange. There have been some successful practices of \textit{science communication on social media}, such as expert-written explanatory tweets~\cite{gero2021makes} and science-centered videos~\cite{xia2022millions,zhang2023understanding}.

Effective science communication has increasingly put an emphasis on the participatory practice with public engagement, rather than one-way information transmission, to facilitate the knowledge construction and comprehension among the public~\cite{miller2001public,chilvers2010sustainable}. To this end, \textit{public making sense of research through social media} is an increasingly popular and crucial process in effective and engaging science communication. Users gather to discuss and develop an understanding of research in general social media sites such as Twitter~\cite{davies2016science,lopez2018social}, science-centered communities such as r/science~\cite{jones2019r} and r/AskHistorians~\cite{gilbert2020run}, and comment threads of scientific articles~\cite{williams2021effects}. With a wide user base mixing domain experts and the general public, people could broadly exchange information~\cite{jones2019r} and learn from other users' opinions~\cite{williams2021effects} to conceptualize the research and draw implications from it to guide daily life (e.g., the wide COVID-19-related e-print discussion~\cite{yudhoatmojo2021we}).

In particular, Community Question and Answering (CQA) platforms afford a valuable channel for public sensemaking of specific research with focus questions. In this setting, \textit{askers} propose questions regarding particular research such as its practical implications or method details, \textit{answerers} comprising domain experts and the general public voluntarily share their interpretations with domain expertise and personal opinions, and \textit{viewers} manage to develop their understanding from such QA-based knowledge construction~\cite{liu2015research}. We define such focus questions regarding specific research\footnote{In this study, the scope of research includes both hard science (e.g., biology and physics) and soft science (e.g., social science).} as ``\textbf{research-sensemaking questions}'', under which users could participate in the research-centered discussion and establish their understanding of the work. Such QA-based discussions naturally contextualize the focus problem with curated questions and descriptions, and facilitate knowledge exchange among askers, answerers and viewers~\cite{jin2013users}. Besides, with question-centered discussions less constrained to social networks and communities, users in different domains and with diverse levels of expertise could contribute to and benefit from QA-based science knowledge exchange~\cite{ZhihuSciComm}. On Zhihu, one of the largest CQA platforms in China~\cite{ZhihuPlatform}, some research-sensemaking questions received thousands of answers and upvotes and millions of views~\cite{ZhihuSciComm}.

As the component initiating discussion, the art (and science) of crafting questions in CQA platforms is significant to elicit high-quality discourse~\cite{ravi2014great}. An engaging research-sensemaking question plays a crucial role in attracting public participation and promoting effective knowledge construction~\cite{liu2015research,liang2019scientists}. Moreover, the establishment of research-sensemaking questions might inherit critical challenges of science communication that hinder engaging a broad audience or ensuring the discussion quality~\cite{august2020explain,williams2022hci}, which further necessitates strategic and careful questioning. On the one hand, research-centered discussions typically include specialized linguistic use (e.g., terminologies and hedge words such as ``likely'' or ``might''), which may become barriers to public participation~\cite{august2020explain}. On the other hand, loosing the constraints on rigor and accuracy to involve a wide audience inevitably brings misuse and misinterpretations of research outputs, such as exploiting science work for conspiracy theories and extremist ideology~\cite{yudhoatmojo2021we,lee2021viral}. To this end, \textbf{how to strategically propose research-sensemaking questions}, considering both accessibility and accuracy for public participation, is a significant yet challenging problem. Nonetheless, which strategies are naturally developed in these research-sensemaking questions to engage wide and high-quality responses, and how they correlate with user participation and knowledge co-construction, are still less understood. Meanwhile, how collaborative question editing, an important feature of CQA platforms in enhancing question quality~\cite{chen2020changing}, influences the construction of research-sensemaking questions is also underinvestigated. Therefore, we propose the following research questions:

\begin{itemize}

  \item \textbf{RQ1}: What are users' strategies in proposing research-sensemaking questions in CQA platforms?
  
  \item \textbf{RQ2}: How do users' strategies in proposing research-sensemaking questions correlate with public engagement and knowledge construction in CQA platforms?

  \item \textbf{RQ3}: How do users strategically apply collaborative editing to improve research-sensemaking questions in CQA platforms?

\end{itemize}

To answer these questions, we collected 837 science-related questions with 157,684 answers from the Zhihu platform, and conducted a mixed-methods study to investigate the questioning strategies and their potential effects. Through an open coding approach (RQ1), we captured users' linguistic and non-linguistic strategies in curating research-sensemaking question titles and descriptions, which reflected users' efforts for both rigorous (e.g., hedging and data use) and attractive (e.g., counter-intuitive and emotional statements) questioning. By quantifying knowledge construction in answers with text classifiers and applying regression analysis (RQ2), we systematically uncovered how these strategies correlated with public participation and knowledge construction, especially the divergence between engaging a wider audience and attracting epistemic and argumentative answers. Through inductive log analysis (RQ3), we revealed the value of collaborative work in constructing and refining research-sensemaking questions, such as co-establishing the topic scope and scientific reframing for rigorous presentations. Based on the findings, we discuss design implications for accessible, accurate, and effective research sensemaking in CQA platforms.

This work makes the following contributions to science communication in HCI and CSCW communities: (1) we deepened the understanding of questioning as an emerging pattern of user participation in science communication, and captured a comprehensive taxonomy of user-developed strategies in proposing engaging and rigorous research-sensemaking questions; (2) we revealed how different questioning strategies correlated with user engagement and knowledge construction, and unpacked the tension between wide participation and quality knowledge establishment; (3) we demonstrated the opportunities and challenges of collaborative editing in crafting research-sensemaking questions. With the trend of involving, engaging and empowering the public in research-related discussions, this work provides rich dynamics on attracting user participation and improving knowledge construction that may shed light on effective science communication. 

\section{RELATED WORK}

This section contextualizes the current study within HCI and CSCW literature on science communication and CQA platforms. We first present an overview of science communication on social media in Section \ref{RelatedWork-SciComm}, outlining its existing practices and challenges on the participatory web. Under the challenge of engaging users with scientific knowledge, we survey the existing literature on communication strategies for science communication in Section \ref{RelatedWork-SciCommStrategy}. We finally situate science communication within CQA platforms in Section \ref{RelatedWork-SciCommCQA}, navigating how the sociotechnical context of CQA platforms may shape the science communication practice.

\subsection{Science Communication on Social Media}\label{RelatedWork-SciComm}

Science communication refers to \textit{``the use of appropriate skills, media, activities, and dialogue to produce one or more of the following personal responses to science: Awareness, Enjoyment, Interest, Opinion-forming, and Understanding''}~\cite{burns2003science}. With the development of social media, science communication has gradually evolved from one-way information dissemination to two-way participatory interactions~\cite{jones2019r,williams2022hci}, which focus more on cultivating public awareness of science through dialogue~\cite{miller2001public} and public engagement~\cite{chilvers2010sustainable}. Scholars in HCI and CSCW have begun to investigate how different social media platforms afford science communication, such as general social media sites (e.g., Twitter)~\cite{gero2021makes,gruzd2020coding}, online communities (e.g., Reddit)~\cite{jones2019r,august2020explain,gruzd2020coding}, and video sharing platforms (e.g., Youtube)~\cite{xia2022millions,zhang2023understanding}.

Characterized by two-way participatory interactions, diverse \textit{practices} of science communication emerged on social media, informing and engaging the public with science knowledge. For example, \textit{publicizing one's own research} has been widely adopted by researchers on social media sites like Twitter~\cite{gero2021makes,keng2023researchers}. By sharing their recent or ongoing work, researchers exchange knowledge with peers and colleagues as well as establish academic connections, which benefit them in developing research~\cite{darling2013role,williams2022hci}. This process may also promote the impact of work by attracting users' attention to the research progress with comprehensible explanations (e.g., well-crafted explanatory Twitter threads as ``tweetorials'')~\cite{gero2021makes}. \textit{Science popularization} on social media, the summarization, simplification, and interpretation of specific scientific topics, has also become a common practice such as scientific blogs~\cite{gardiner2018you,luzon2013public} and videos~\cite{zhang2023understanding,xia2022millions}. The participatory interactions, such as user-generated comments, make science popularization on social media more accessible when they help to facilitate others' understanding and establish the feedback mechanism on the content quality~\cite{he2021beyond,zhang2023understanding}. Apart from science communication with a central communicator, users also gather on social media to \textit{collectively make sense of research}. For example, online science communities (e.g., r/science~\cite{jones2019r,august2020explain} and r/AskHistorians~\cite{gilbert2020run,gruzd2020coding}) attract group members who share similar interests yet have diverse levels or focuses of expertise, which provide an effective channel of peer learning~\cite{jones2019r} and knowledge co-construction~\cite{jones2019r,he2022help}.

On the other hand, science communication on social media also suffers from various \textit{challenges}, including linguistic barriers that hinder reaching and engaging audiences with less domain knowledge~\cite{kopke2019stepping,bruine2013assessing,august2020explain}, difficulties in explaining complicated topics to laypeople accurately~\cite{rice2017contexts,ecklund2012academic}, the potential risks of misinformation and misinterpretation~\cite{lee2021viral,cook2018benefits,yudhoatmojo2021we}, and the danger of harassment to communicators~\cite{stewart2016collapsed}.

Given the opportunities and challenges of science communication on social media, it is important to understand user participation and engagement in the specific sociotechnical context, and design corresponding interfaces to support them. Nonetheless, the existing literature lying in the intersection between science communication and HCI is still scarce. There is limited understanding of QA-based science communication, especially in a non-western context~\cite{williams2022hci}. This work aims to contribute to this venue by unpacking strategies of proposing researching-sensemaking questions and their potential effects in a Chinese CQA platform.

\subsection{Science Communication Strategies}\label{RelatedWork-SciCommStrategy}

Given the importance and difficulties in communicating engaging, understandable, and accurate scientific knowledge~\cite{rice2017contexts,williams2022hci}, effective communication strategies are crucial components in science communication on social media. In this section, we review the literature on science communication strategies from the perspective of \textit{information} and \textit{interaction}.

How to craft good science-related \textit{information} for the general public, whether textual or visual, has been widely investigated by scholars in education and communication~\cite{kendall2007write,blum2005field,TipsForSciComm,finkler2019power,gero2021makes}. For example, Dahlstrom argued that using narratives and storytelling to communicate science with nonexpert audiences may enhance comprehension, interest, and engagement~\cite{dahlstrom2014using}. Flemming et al. revealed the value of emotionalization in science communication such as promoting knowledge gain~\cite{flemming2018emotionalization}. Researchers also concluded knowledge-crafting strategies in more specific settings. For science writing, Gero et al. summarized some good practices including \textit{an implicit structure, an attention-grabbing lede, specific stories as a driving force, explanatory strategies like analogy and metaphor}, and \textit{return-to-question conclusions}, and further explored their use on scientific Twitter threads created by experts~\cite{gero2021makes}. For video-based science communication, Finkler and Le{\'o}n proposed a SUCCESS framework of the visual rhetoric that urged for producing videos that are Simple, Unexpected, Concrete, Credible, Emotional, and Science Storytelling~\cite{finkler2019power}.

A growing body of work in science communication has also turned to the \textit{interactions} among communicators and audiences, beyond building high-quality scientific information, that may contribute to effective science communication. One line of work explored enhancing and deepening scientist-public dialogue to promote science communication~\cite{peters2013gap, chilvers2010sustainable,falchetti2007laypersons,zorn2012influence}. For example, Zorn et al. found that dialogue between laypersons and scientists on human biotechnology could make scientists' and laypeople's attitudes toward human biotechnology converge, and increase laypeople's communicative self-efficacy~\cite{zorn2012influence}. Another strand of work focused on involving the public to participate in and contribute to science communication, taking the public as not merely information receivers but also knowledge contributors~\cite{jones2019r,pandey2017gut,williams2021effects,he2021beyond,tinati2015designing}. This process could exploit collective wisdom in constructing and making sense of scientific knowledge through opinion sharing, knowledge exchange, discussion and debates~\cite{jones2019r}. For instance, users' comments on science news or videos not only reflect the public understanding of the transmitted knowledge~\cite{dubovi2021interactions}, but also influence others' perceptions and consumption of the scientific information~\cite{williams2021effects}.

This work contributes to the understanding of science communication strategies from both \textit{information} and \textit{interaction} perspectives. For \textit{information}, we unveiled users' linguistic and non-linguistic strategies in asking and describing researching-sensemaking questions, and how they correlated with public participation and knowledge construction. For \textit{interaction}, we investigated how collaborative work was presented in CQA platforms and facilitated science communication, including collaborative editing in questions and knowledge co-construction in answers.

\subsection{Community Question Answering (CQA) Platforms and Science Communication}\label{RelatedWork-SciCommCQA}

Community Question Answering (CQA) platforms, such as Quora, Stack Exchange, and Zhihu, have been crucial online information hubs that millions of users turn to for information seeking and sharing~\cite{srba2016comprehensive}. CQA platforms attract users with different experiences and expertise to participate in the QA-based discussions, and largely benefit from the ``wisdom of crowds'' in knowledge construction and exchange~\cite{srba2016comprehensive,wang2013wisdom}. 

As a typical and important CSCW system~\cite{hadi2022users}, CQA platforms have gained research attention in HCI and CSCW communities from many different perspectives. Some scholars evaluated \textit{user behaviors} such as answering~\cite{yang2014sparrows,pal2012evolution,anderson2012discovering} and lurking~\cite{khansa2015understanding} on CQA platforms. Generally, users' preferences in choosing questions to answer were rather different~\cite{yang2014sparrows}, and only a few answerers contributed a large share of answers~\cite{nam2009questions}. Different users had substantially distinct activity patterns (e.g., questioning, answering, and socializing) and linguistic features, making it easy to automatically identify experts and non-experts on CQA platforms~\cite{patil2016detecting}. Some prior work examined the factors that influenced \textit{information quality} and the effects of information quality on user behaviors on CQA platforms, covering both questions~\cite{ravi2014great,wang2013wisdom,asaduzzaman2013answering} and answers~\cite{harper2008predictors,li2015answer}. For example, Ravi et al. found that both the question topic and length were significant predictors of question quality in CQA sites~\cite{ravi2014great}. Asaduzzaman et al. revealed a set of features from unanswered questions that potentially made them hard to get answers, such as those too specific or hard to follow~\cite{asaduzzaman2013answering}.

Another line of work focused on the dynamics of \textit{user identity} in CQA platforms~\cite{vargo2018identity,guo2021anonymity,dubois2022towards,das2021jol}, which largely shapes the ecosystem of Q\&A communities and influences user engagement under the sociotechnical context. For example, Das et al. revealed that the sociotechnical mechanisms of governance on Bengali Quora would privilege certain identities and marginalize others regarding linguistic practices, nationalities, and religious affiliations~\cite{das2021jol}. In the same vein, Dubois et al. found that Q\&A community cultures contributed to gender differences in contribution styles and user appreciation~\cite{dubois2020gender}, and Gilbert noted that the default masculine whiteness of Reddit challenged the moderation of Q\&A in a history-centered academic sub-community~\cite{gilbert2020run}. More recent work also explored the influences of \textit{platform-specific features} on user participation and collaboration, such as the affordances of co-editing questions~\cite{chen2020changing} and flagging mechanisms~\cite{li2022flagging}.


Science communication on CQA platforms shares some similarities with other community-based science communication, and also has its unique characteristics. First, the QA-initiated science discussion on CQA platforms is similar to science-centered QA communities such as r/AskHistorians~\cite{gilbert2020run}, which necessitates both high-quality questions and answers for effective science communication. Such question-answering activities extend beyond one-off answers to a knowledge co-creation process, developing long-lasting value~\cite{anderson2012discovering}. Also, the science knowledge co-construction in public-generated answers is a typical example of argumentative knowledge construction in computer-supported collaborative learning (CSCL)~\cite{weinberger2006framework}. Therefore, this work followed the framework proposed by Weinberger and Fischer to analyze user participation and quality knowledge construction in answers, including dimensions of \textit{participation, epistemic, argumentative}, and \textit{social mode}~\cite{weinberger2006framework}. 

On the other hand, compared to science-related subreddits with community members as the major participators~\cite{gilbert2020run,jones2019r,august2020explain}, science communication on CQA platforms is more question-oriented and less constrained by the community size, focus, and norms, thus demanding careful curation of questions to make them attractive and accessible to wider users. That necessitates the understanding of well-crafted question-asking strategies to attract and engage the general public, which is the focus of this study. Relevant to it, Zhihu allows collaborative editing of questions to improve the question quality and make them applicable to a broader audience~\cite{chen2020changing}. We also examined such collaborative work in enhancing the quality of research-sensemaking questions.

\section{METHOD}

\begin{figure}
	\centering
		\includegraphics[scale=.37]{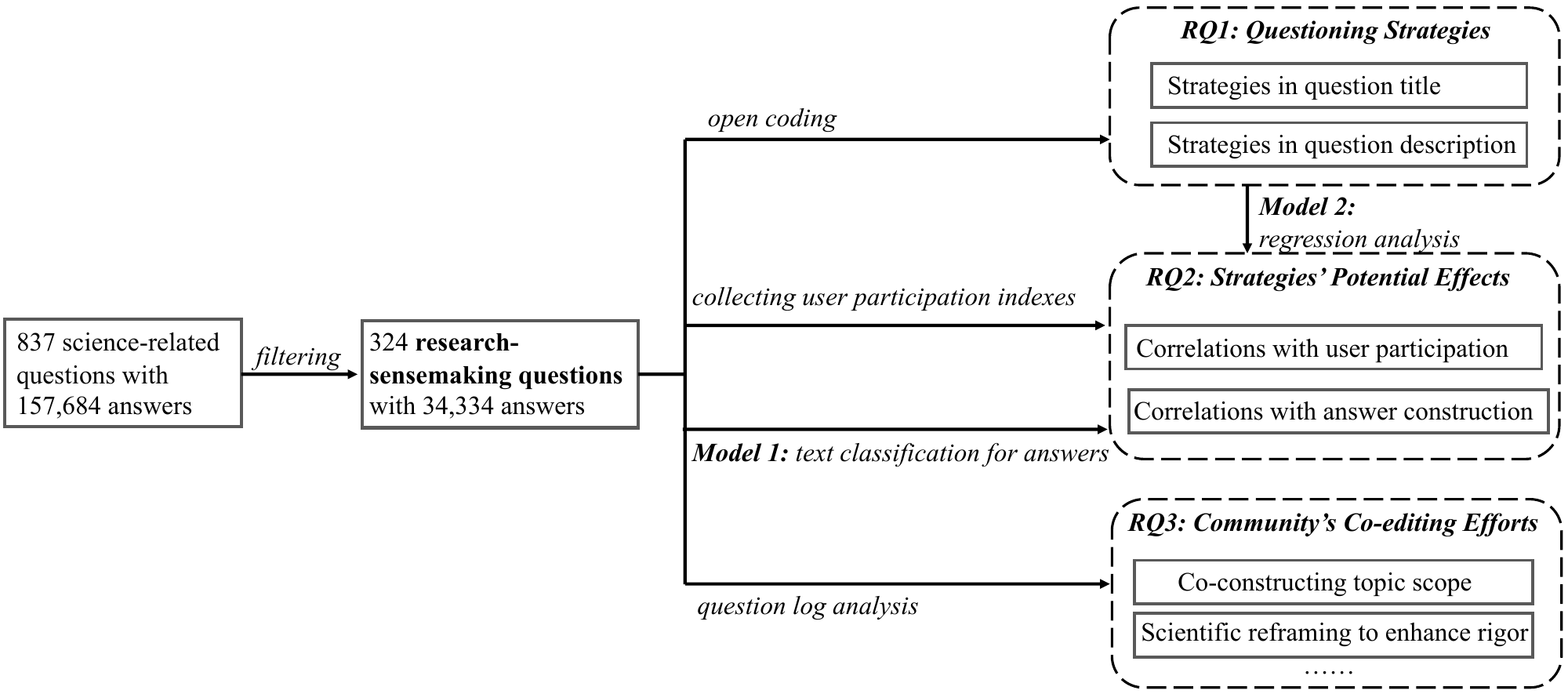}
	\caption{The analytical flow to understand user-developed strategies in proposing research-sensemaking questions.}
	\label{FIG: method}
\end{figure}

This section describes the mixed-methods approach used to understand the questioning strategies for research-sensemaking on a CQA platform. We first briefly introduce the platform in Section \ref{platform}, the data collection process in Section \ref{dataCollection}, and the dataset description in Section \ref{dataDes}. Then, we present (1) how we adopted \textit{open coding} to capture strategies in proposing research-sensemaking questions (RQ1, Section \ref{RQ1:method}); (2) how we quantified quality knowledge construction with \textit{text classification}, and applied \textit{regression analysis} to figure out questioning strategies' correlations with user engagement and answer development (RQ2, Section \ref{RQ2:method}); and (3) how we conducted \textit{log analysis} to unpack the community's efforts to craft research-sensemaking questions through collaborative editing (RQ3, Section \ref{RQ3:method}). The overall analytical flow is shown in Figure \ref{FIG: method}.

\subsection{Platform}\label{platform}

\begin{figure}
	\centering
		\includegraphics[scale=.35]{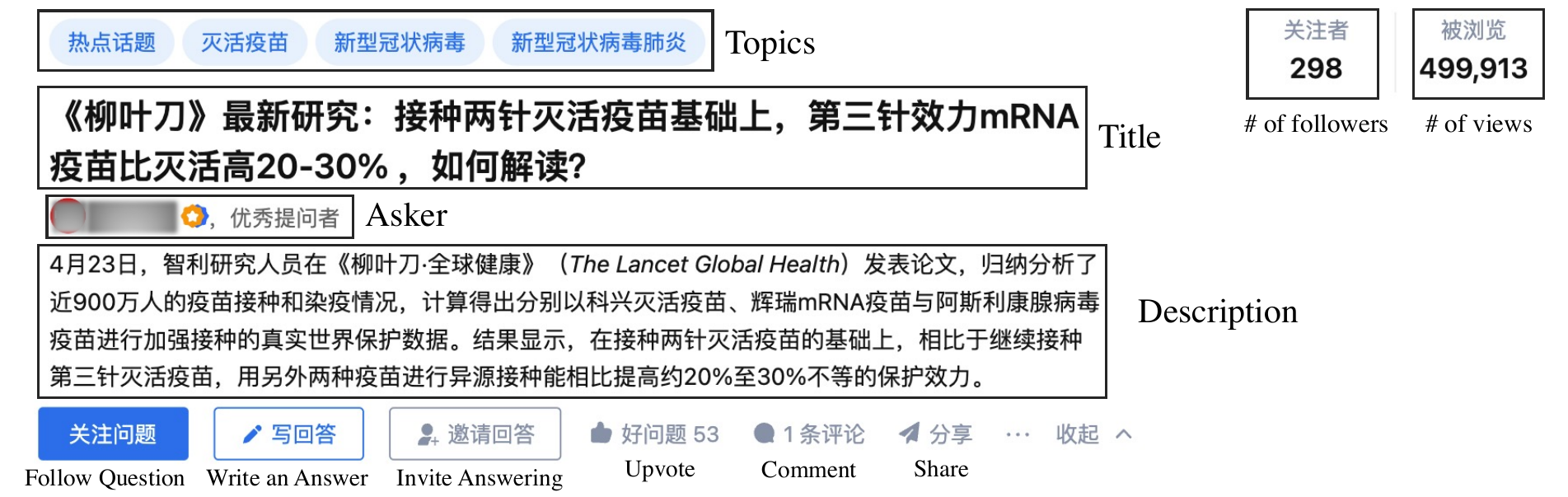}
	\caption{An example of Zhihu question interface. The example research-sensemaking question is ``\textit{The latest research in The Lancet: With the basis of two doses of inactivated vaccine, the efficacy of a third dose of mRNA vaccine is 20-30\% higher than that of inactivated vaccine. How to interpret it?}''}
	\label{FIG: example}
\end{figure}

We situated this study on Zhihu, one of the most popular CQA platforms in China. As of May 2022, Zhihu had more than 100 million average monthly active users~\cite{ZhihuUsers} and accumulated more than 44 million questions~\cite{ZhihuQuestions}. The basic interfaces of Zhihu are similar to Quora, such as the affordances of question asking, question answering, question following, and question commenting. However, different from Quora which had disabled the ``question details'' feature~\cite{QuoraDescription}, Zhihu allows users to provide question descriptions with formatted text (such as highlighted quotes, code blocks, and formulas), links, figures, and videos~\cite{guo2021anonymity}. This feature is important for users to supplement research-sensemaking questions, typically complex in nature, with detailed explanations of research and specific questions. Besides, Zhihu also affords comprehensive features for collaborative question editing from modifying question titles and question descriptions to adding or removing question topics~\cite{chen2020changing}. This interface supports collaborative work to enhance the quality of research-sensemaking questions. An example of question-related interfaces of Zhihu is shown in Figure ~\ref{FIG: example}.

\subsection{Data Collection}\label{dataCollection}

We first conducted an exploratory scanning step to determine the data inclusion criteria for research-sensemaking questions. Our goal was to generate a filtering approach for research-sensemaking questions that were (1) unbiased in science branches, (2) inclusive to capture different ways of asking research-sensemaking questions, and (3) able to largely exclude irrelevant data. Therefore, we did not choose topic-based collection that utilized science-related topics as hashtags to collect data, as we found that many research-sensemaking questions did not have subject hashtags, and science-related topics had miscellaneous question types (e.g., the ``physics'' topic had many general questions similar to ``how great is Einstein?'' or ``What is the longest formula?''). Instead, we chose keyword-based collection, and carefully designed the keyword set. To do so, we first randomly picked ten research-sensemaking questions that appeared on the ``hot'' page (top 50 popular questions in real-time) on the Zhihu platform in 2022. Taking these questions as the seeds, we continuously went through ``related questions'' recommended by Zhihu to examine similar research-sensemaking questions, through which process we managed to summarize some general patterns that research-sensemaking questions were asked. Before adding one keyword to the keyword set, we manually checked the search results for impartiality, diversity, and purity (e.g., the keywords biased toward specific disciplines were excluded). After these steps, we reached a final keyword set to collect research-sensemaking questions, including: \textit{``research finds'', ``research claims'', ``research shows'', ``research proves'', ``research indicates''}, and \textit{``research reveals''}. These keywords characterized research-sensemaking questions yet were not skewed towards specific questioning strategies or science branches.

Taking the refined keywords as the search queries, we collected all questions that appeared in the search results through web scraping. Specifically, we searched every keyword on Zhihu without logging into an account to mitigate the influence of personalization~\cite{hussein2020measuring}. We customized the search by limiting the search results to only questions (i.e., excluding articles and videos), adopting the default relevance ranking of Zhihu, and setting the period to ``anytime''. The Zhihu search engine returned 100 to 200 most relevant questions for each search. Then, we used the question IDs crawled from the search page to find the question page so that we managed to collect the data of each question as well as its answers. Note that due to the finite size of returned search results on Zhihu, we could not establish a comprehensive dataset. However, the curation of the keyword set and the collection in an anonymous search setting ensured the representativeness and randomness of collected research-sensemaking questions, which constructed the basis for capturing questioning strategies.

We completed the data collection in October 2022, generating an initial dataset including 837 science-related questions with 157,684 answers. We then reviewed each question to filter out those non-research-sensemaking questions, following the criteria of (1) whether it focused on specific research papers or projects; (2) whether it raised specific questions about the scientific research. It narrowed down the dataset to 324 distinct questions with 34,334 answers, covering a 10-year range from January 2012 to September 2022. The collected metadata of questions included \textit{question id, title, link, created time}, and the number of \textit{answers, views, followers, comments} and \textit{upvotes}. We further collected asker-related information including \textit{anonymity, follower number} and \textit{following number}. The metadata of answers included \textit{answer id, answer content, created time, author id}, and the number of \textit{thanks, comments} and \textit{upvotes}.

\subsection{Dataset Description}\label{dataDes}

The preprocessed dataset contained 324 distinct research-sensemaking questions with 34,334 answers. The questions covered science branches including \textit{health \& medical science} (43.8\%), \textit{life science} (23.8\%), \textit{earth science \& space science} (13.0\%), \textit{social science \& arts} (12.7\%), \textit{math \& physics} (4.3\%), \textit{computers \& technology} (1.2\%), and \textit{chemistry \& material science} (1.2\%). A proportion of 30.9\% questions aimed to make sense of research topics that were strongly \textit{decision-related} (e.g., ``\textit{Whether to get COVID-19 vaccine boosters}''), and the remaining were less close to personal decision-making (e.g., ``\textit{Why the dinosaurs became extinct}'').

In addition to general sensemaking questions typically structured as ``\textit{How do you view/evaluate/think of/understand/interpret (the research)}'', we identified some extended \textit{sensemaking goals}, including: (1) sensemaking and discussing implications (49.7\%), e.g., ``\textit{Research found that `exercise actually has a very limited effect on weight loss', so is it still meaningful to exercise to lose weight? How to lose weight scientifically?}''; (2) sensemaking and credibility assessment (9.9\%), e.g., ``\textit{Research claims that the HPV vaccine may lead to increased female infertility. Is it true?}''; (3) sensemaking and reasoning (5.6\%), e.g., ``\textit{How do you view the Chinese social mentality research report stating that `Chinese men have a stronger sense of fairness than women'? What might be the reasons?}''

The questions had an average \textit{title length} of 37.8 Chinese characters (SD=8.3) and an average \textit{topic (tag) size} of 4.5 (SD=0.9). A proportion of 68.2\% questions were asked by non-anonymous users.

The aforementioned dimensions naturally characterized the questions and might intrinsically influence how users were engaged in viewing and answering (e.g., text length and hashtag numbers are potential factors that influence user engagement on social media~\cite{gkikas2022text,de2020sponsored}). Therefore, we took these dimensions as control variables in regression analysis to be demonstrated in Section \ref{RQ2-Findings}.

\subsection{RQ1: Identifying Users' Strategies in Proposing Research-sensemaking Questions}\label{RQ1:method}

We took an open coding approach~\cite{flick2022introduction} to inductively identify how users strategically propose research-sensemaking questions, letting the codes naturally emerge from the analysis. Specifically, two authors independently coded 50 initial samples of research-sensemaking questions, focusing on both linguistic and non-linguistic features that might enhance science communication and attract user engagement. In particular, the analysis examined two separate parts, including the \textit{question title}, the brief question text that introduced the research and the focus of sensemaking; and the \textit{question description}, which allowed detailed explanations of the research for sensemaking in formatted text, visuals, and hyperlinks. After generating initial codes, two coders took several rounds of meetings, comparisons and discussions to reach a consensus on the codebook. Example codes were ``\textit{data use}'' and ``\textit{counter-intuitive statements}''. Finally, the two coders applied affinity diagramming~\cite{corbin2014basics} to organize the codes and group similar codes into high-level themes. Example themes were ``\textit{significance signs}'' and ``\textit{eyecatching narratives}''.

To provide a quantitative description of strategies as well as prepare for regression analysis, the two coders took an annotation round on all 324 research-sensemaking questions based on the codebook, i.e., evaluating each question on whether it applied these strategies. They first re-coded the 50 initial samples independently and compared their labels on each dimension to test the inter-rater reliability. With the agreement ratio (accuracy) greater than 0.9 and Cohen's Kappa~\cite{mchugh2012interrater} greater than 0.8 for every dimension, substantial agreement was achieved between the two coders. Finally, they coded half of the remaining 274 questions separately, yielding 324 fully-labeled research-sensemaking questions. We took a similar approach to identify other basic features of questions (i.e., \textit{science topic} and \textit{sensemaking goal} in Section \ref{dataDes}) as control variables for regression, which fundamentally characterized these questions but did not reflect strategic question construction.

\subsection{RQ2: Investigating Questioning Strategies' Potential Effects}\label{RQ2:method}

RQ1 uncovered a comprehensive taxonomy of user-developed strategies in asking and describing research-sensemaking questions on Zhihu. In this section, we further investigated how these research-questioning strategies correlated with user participation and knowledge co-construction through regression analysis. Specifically, we quantified \textit{user participation} with public engagement indexes of questions (\textit{views}, \textit{upvotes}, \textit{followers}, and \textit{answers}), and used \textit{Poisson regression} to predict these count dependent variables. We examined quality \textit{knowledge construction} based on Weinberger and Fischer's framework of argumentative knowledge construction in CSCL~\cite{weinberger2006framework} (i.e., quantifying proportions of \textit{epistemic, argumentative}, and \textit{social} answers for each question with text classification), and applied \textit{Beta regression} to investigate questioning strategies' correlations with these proportion dependent variables.

\subsubsection{Dependent Variables: Measuring User Participation}

For each research-sensemaking question, we took public engagement indexes (\textit{views, upvotes, followers,} and \textit{answers}) as the reflection of the general \textit{user participation}. They measured the \textit{quantity} of users engaged in knowledge construction~\cite{weinberger2006framework} and research sensemaking, as described below.

\begin{itemize}
  \item \textbf{Participation}
    \begin{itemize}
    
        \item \textit{Views} (count variable): The number of users viewing the question. It reflected users' general attention to the research-sensemaking question.
        
        \item \textit{Upvotes} (count variable): The number of users upvoting the question, which action would make the question be recommended to more users. It captured users' approval of the quality of the research-sensemaking question.

        \item \textit{Followers} (count variable): The number of users following the question, which action would make the users get notified about new answers for the question. It represented users' long-term interest in the research-sensemaking question.

        \item \textit{Answers} (count variable): The number of users answering the question. It showed users' contributions to the research-sensemaking question.
        
    \end{itemize}
\end{itemize}

\subsubsection{Dependent Variables: Quantifying Knowledge Co-construction Based on Text Classification}

We built various types of binary text classifiers for answers regarding \textit{epistemic, argumentative}, and \textit{social} dimensions~\cite{weinberger2006framework}, which represented the \textit{quality} of knowledge co-construction from different aspects, i.e., whether the answer was on-topic, well-justified, and taking others' thoughts into consideration. For each research-sensemaking question, we calculated the proportions of answers having these specific features to measure the knowledge co-construction dimensions.

\begin{itemize}
  
  \item \textbf{Epistemic}: The epistemic dimension estimates how people work on the knowledge construction task they face~\cite{fischer2002fostering}, and the primary task is to examine whether users are engaging in activities to solve the task (\textit{on-task discourse}) or rather concerned with off-task aspects~\cite{weinberger2006framework}. Specific to the scenario of QA-based research sensemaking, we defined such \textit{on-task discourse} as on-topic answers that contributed to the research-sensemaking question in knowledge, in contrast to off-topic digressions (e.g., just venting emotions to the research/researcher). Therefore, we applied the following dependent variable to measure the \textit{epistemic} dimension for each question:
  \begin{itemize}
    \item \textit{On-task discourse} (continuous variable): The proportion of on-topic answers contributing to research-sensemaking in knowledge.
  \end{itemize}
  
  To measure this variable, we trained a pairwise text classification model to identify the semantic relations for question-answer pairs (on-topic vs. off-topic). The specific process was in four steps: (1) \textit{annotation}. we randomly sampled 1000 question-answer pairs from the whole dataset. Two authors independently coded the first 100 samples to determine whether the answer contributed to the research-sensemaking question in knowledge, and assigned the on-topic or off-topic label. The Cohen's Kappa reached 0.85 (accuracy=94\%), indicating substantial agreements between coders. After several rounds of discussions to resolve the difference, the two authors further annotated 450 question-answer pairs each, generating 1000 label-assigned samples. (2) \textit{building a pairwise text classifier for on-task discourse}. We performed the text classification using Bidirectional Encoder Representations from Transformers (BERT). We used the BERT model not only for its good performance across different NLP tasks, but also considering that the base training task of next sentence prediction in BERT supported the fine-tuning for sequence pair classification~\cite{devlin2018bert}. We adopted the Chinese pretrained model of BERT-wwm~\cite{cui2021pre}. We used the basic structure of sequence pair classification~\cite{ostendorff2020pairwise}, i.e., separating tokens from questions and answers with the [SEP] token, identifying the two types with a binary mask (\textit{token\_type\_ids}), and jointly feeding them through the model. Taking 900 samples as the training set and 100 samples as the test set, the text classification achieved good performance with the F1-score = 87.7\% on the test set. (3) \textit{predicting on-task discourse for question-answer pairs}. We applied the classifier to assign the label of \textit{on-task discourse} for all 34,334 question-answer pairs. (4) \textit{quantifying epistemic dimension for questions}. For each question, we calculated the percentage of answers with on-task discourse as the index of epistemic dimension (ranging from 0 to 1). 

  \item \textbf{Argumentative}: The argumentative dimension measures the construction of arguments facing complex problems~\cite{weinberger2006framework}. We took a simplified framework of argumentative claims to evaluate whether and how answers were argumentative~\cite{mcneill2011claims}, i.e., \textit{evidence} and \textit{reasoning}.

  \begin{itemize}
    \item \textit{Evidence} (continuous variable): The proportion of answers that provided explicit evidence that supported the claim, such as statistical data and theories.
    \item \textit{Reasoning} (continuous variable): The proportion of answers that logically justified why the claim was valid.
  \end{itemize}

  We also took a text classification approach to measure the two argumentative indexes. The process was similar to measuring the Epistemic dimension, except that the task became single-document classification for answers. Steps included (1) initial coding (N=100, with Cohen's Kappa = 0.86 for \textit{evidence} and Cohen's Kappa = 0.81 for \textit{reasoning}), discussing, and disagreement resolving; (2) two coders' annotation to generate a label-assigned dataset (N=1000); (3) building BERT-based classifiers based on 900 training samples and evaluating on a test dataset of 100 samples (F1 score = 0.86 for \textit{evidence}, F1 score = 0.84 for \textit{reasoning}); (4) prediction to scale up the labels to the whole answer dataset; and (5) calculating the proportion of answers with \textit{evidence} and \textit{reasoning} labels as the measurement of argumentative dimension for each research-sensemaking question. 
  \item \textbf{Social} (continuous variable): Users might socialize with others to enhance knowledge co-construction, which is a crucial component in CSCL for knowledge acquisition and establishment~\cite{weinberger2006framework}. Typical social modes in our dataset included (1) elicitation, e.g., ``\textit{According to my understanding, the logic seems to be correct? (some details) I hope some experts can explain the flaws of the article}''; (2) agreeing and supplementing, e.g., ``\textit{The main contributions of this research have already been explained by @[User]. For details, please refer to this answer: [LINK]. But I must make some elaborations on some places that are easily overlooked...}''; (3) disagreeing and rebutting, e.g., ``\textit{I object to @[User1] @[User2]'s claim that the population of the United States will continue to grow steadily...}'' Considering all these potential social modes, we measured the social dimension as:
  \begin{itemize}
    \item \textit{Social} (continuous variable): The proportion of answers socializing with other users for knowledge exchange, discussion and establishment.
  \end{itemize}
  
  We followed the same text classification-based approach as the \textit{evidence} and \textit{reasoning} indexes to compute the \textit{social} proportion for each question (Cohen's Kappa = 0.90 during the first round of coding). Due to the lower frequency compared to other dimensions, we coded a larger training set (N=1500), and evaluated 100 positive samples (recall=0.89) after prediction to ensure its practicality.

\end{itemize}

\subsubsection{Regression Analysis}

We took questioning strategies identified in Section \ref{RQ1:method} as the independent variables to explore how they correlated with the various dimensions of user engagement and knowledge co-construction. Considering the influence of basic question characteristics, we also included control variables covering: (1) the topic features, i.e., science branches and relevance to decision making; (2) sensemaking goals; (3) asker information, i.e., anonymity, follower and following; (4) title length and topic (tag) size, as shown in Section \ref{dataDes}. We used only question title-related strategies to predict the count of \textit{views} as users could not see the description before opening the question, and combined title and description strategies in predicting the remaining indexes. We pre-processed the independent variables including one-hot encoding for categorical variables (e.g., science branches of the topic and goals of the question) and normalization.

We applied (1) Poisson regression for count dependent variables (i.e., \textit{views, upvotes, followers} and \textit{answers}), which could effectively estimate count data~\cite{coxe2009analysis}; and (2) Beta regression for continuous dependent variables as proportions (i.e., \textit{on-task discourse, evidence, reasoning}, and \textit{social}), which is well-suited to dependent variable in the form of fractions or percentages~\cite{ferrari2004beta}. As highly related features would lead to poor estimation in regression, we tested multicollinearity in features through Variance Inflation Factor (VIF)~\cite{akinwande2015variance}. We found all features had VIF < 3.5 (VIF of 5 and above generally suggests high multicollinearity and bad regression performance~\cite{akinwande2015variance}), indicating low multicollinearity. We excluded questions with fewer than 5 answers to reduce the small-sample bias in regressions for continuous dependent variables that used proportion-based estimation. We excluded new questions asked within 1 month before the collecting date in regressions for count dependent variables as their count indexes might not reach stability.

\subsection{RQ3: Understanding Community's Collaborative Work in Constructing Research-sensemaking Questions}\label{RQ3:method}

RQ1 and RQ2 investigated users' strategies in proposing research-sensemaking questions on a CQA platform, and their correlations with user participation and knowledge construction. This section further examined the construction of research-sensemaking questions from the perspective of community's collaborative work, unpacking how community members collectively crafted research-sensemaking questions and improved their quality through collaborative editing.

To achieve this goal, we conducted inductive coding on question logs that recorded users' collaborative editing. Specifically, two authors independently analyzed logs of the questions that enabled collaborative editing, letting the codes naturally emerge from the analysis. For each question, they first read through all editing logs, including the specific edits and reasons for edits, from the created time to the time when the collaborative editing was locked\footnote{Collaborative editing of Zhihu questions would be locked when: (1) The question has received a specific number of high-quality answers; (2) The question has been collected into the ``hot'' page; (3) The question was identified with little room for improvement by moderators~\cite{ZhihuCollaborativeEditing}.}. They grouped edits into title edits, description edits and tag change~\cite{chen2020changing}, and particularly paid attention to \textit{how the edits reflected specific question-asking strategies, and how they might improve the research-sensemaking questions}. Two authors coded till saturation when no new codes emerged, and reached a consensus through several rounds of comparisons and discussions. In total, they coded 66 questions with 536 edit logs.

\section{FINDINGS}

Through a mixed-methods approach, this work enriches the understanding of how users strategically proposed research-sensemaking questions, and the potential effects. In this section, we first present the taxonomy of user-developed strategies in crafting research-sensemaking question titles and descriptions in Section \ref{RQ1findings} (RQ1). We then demonstrate how these strategies correlated with user participation and quality answer construction in Section \ref{RQ2-Findings} (RQ2), which unpacked the divergence between engaging a wide audience and facilitating on-topic and argumentative discourse in answers. We finally reveal the unique values of collaborative editing from the community in constructing research-sensemaking questions in Section \ref{RQ3-findings} (RQ3).

\subsection{RQ1: Strategies in Proposing Research-sensemaking Questions}\label{RQ1findings}

This section presents the taxonomy of users' strategies in asking research-sensemaking questions (Section \ref{StrategyQTitle}) and providing detailed question descriptions (Section \ref{StrategyQDes}) on a Chinese CQA platform. The findings enlighten how users strategically crafted questions to set up for the research-sensemaking tasks and engage the audience for participation.

\subsubsection{Strategic Question Titles}\label{StrategyQTitle}

\begin{table*}
  \small
  \caption{Strategic Question Titles: The codebook of user-developed strategies in asking research-sensemaking questions}
  \label{tab:question_title}
  \begin{tabular}{p{1.6cm}|p{1.8cm}p{3.3cm}p{4.5cm}p{1.2cm}}
    \toprule
    Category & Strategy & Definition & Example & Proportion \\
    \midrule
    \multirow{4}{1.5cm}[-8em]{\textbf{Significance Signs}} & Impact indication & Use linguistic cues to indicate the impact of the work, such as ``great contribution'' and ``significant breakthrough'' & \textit{Scientists have revealed the origin and evolution of jawed vertebrates, which is \ul{an important breakthrough} in the ``from fish to human'' research. What does this imply?} & 12.3\% \\
    
    & Timeliness indication & Use linguistic cues to indicate the timeliness of the work, such as ``the latest work'' and ``a new study'' & \textit{\ul{The latest research} found that when humans saw food, the brain would have a short-term inflammatory response. How do you view this research?} & 21.0\% \\

    & Publication venue & Include the (typically well-known) publication venue of the research paper to indicate the significance of work & \textit{New research \ul{in Nature} finds potential evidence of the oldest animal fossil, dating back 890 million years. what are the implications?} & 14.8\% \\

    & Researcher background & Include the background of researchers such as institutes, labs or leading scientists (that are typically well-known) to indicate the significance of work & \textit{A study by the \ul{University of Copenhagen} showed that global warming may cause each person to lose 58 hours of sleep per year. What is the specific situation? What are their connections?} & 38.6\% \\
     
    \hline

    \multirow{2}{1.5cm}[-4.5em]{\textbf{Rigorous Descriptions}} & Hedging & Soften research statements, such as using words ``possibly'' and ``sometimes'' & \textit{American scientists have found that the air transmission rate of the coronavirus is \ul{possibly} a thousand times higher than transmission through contact surfaces. How does this finding help the epidemic control?} & 17.9\% \\

    & Data use & Use data to present conclusions or explanations of research & \textit{How to evaluate the latest research in Nature Medicine that uses large samples of real data to prove that the mortality risk of Omicron is reduced by \ul{79\%?}} & 27.2\% \\
    
    \hline

    \multirow{3}{1.5cm}[-6em]{\textbf{Eyecatching Narratives}} & Quoting & Use quotation marks to highlight key statements & \textit{South Africa found a new strain of the coronavirus that \ul{"has a large number of mutations, and potentially evades body defenses"}. What is the specific situation? How should humans protect themselves?} & 34.0\% \\

    & Emotional arousal & Trigger positive or negative emotional responses & \textit{What do you think of the research saying that ``COVID-19 may cause diabetes in healthy people''?} (\ul{anxiety}) & 44.8\% \\

    & Counter-intuitive statements & Propose counter-intuitive research findings (potentially exaggerated or misrepresented) & \textit{How to understand the study saying that "mental illness diagnoses are scientifically meaningless"?} & 32.0\% \\

    \bottomrule
  \end{tabular}
\end{table*}

Table \ref{tab:question_title} describes the linguistic strategies of asking research-sensemaking questions along with their definitions, examples and proportions. Generally, users embed rich information in the short text of question titles to serve for both effectively capturing users' eye and accurately presenting the research information, covering themes including:

\begin{itemize}

    \item \textbf{Significance signs}. We noticed that users commonly adopted significance signs to attract users' attention and help establish the credibility of work. The signs could be explicit as \textit{impact indication} and \textit{timeliness indication}, which directly used linguistic cues like ``great contribution'' or ``the latest work'' to point out the significance of research; or implicit in the way of pointing out the \textit{publication venue} and \textit{researcher background}, indicating the research was published in venues with a good reputation or conducted by famous institutes or scientists. Specific examples could be found in the first category of Table \ref{tab:question_title}.

    \item \textbf{Rigorous descriptions}. We observed two prevalent practices in rigorously presenting the research findings, including adopting \textit{hedging} to soften statements with words such as ``possible'' or ``might'', and applying \textit{data use} to accurately describe the research. Specific examples could be found in the second category of Table \ref{tab:question_title}.

    \item \textbf{Eyecatching narratives}. A common eyecatching practice was \textit{quoting key findings} with quotation marks to highlight research statements for sensemaking. We also identified that users tended to craft research claims with \textit{emotional arousal} (such as raising anxiety among the audience) or apply \textit{counter-intuitive statements} to attract users' attention and participation. Specific examples could be found in the third category of Table \ref{tab:question_title}.

\end{itemize}

Note that these strategies may not be equivalent to ``good practice''. For example, using emotional and counter-intuitive narratives may attract users but potentially introduce exaggerations or misrepresentations, which might bring off-topic discussions; using rigorous descriptions may unintentionally lift the linguistic barrier~\cite{august2020explain}, which might exclude some users in discussion. Therefore, we investigated these strategies' associations with both participation and high-quality knowledge co-construction in answers.

\subsubsection{Strategic Question Descriptions}\label{StrategyQDes}

Due to the complexity of research, question descriptions played an important role for askers to supplement and clarify important details in research-sensemaking questions. We found that question askers strategically crafted question descriptions to engage users to participate in answering or facilitate on-topic and high-quality knowledge construction in answers. Figure \ref{FIG: description_strategy} presents a representative example of users' linguistic and non-linguistic strategies in constructing the description. We conclude their definitions below:

\begin{figure}
	\centering
		\includegraphics[scale=.50]{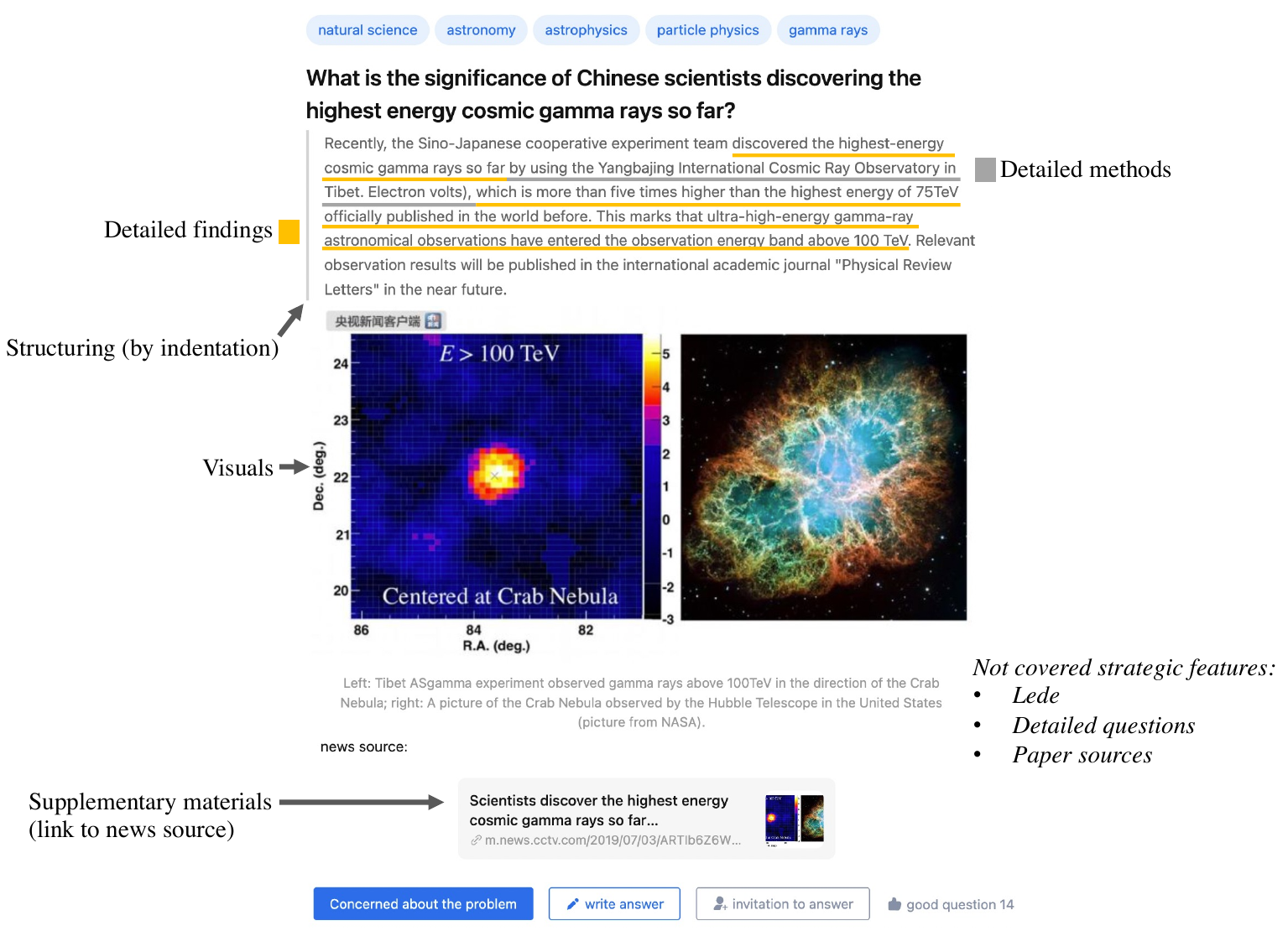}
	\caption{An example of strategies in question descriptions. The screenshot of the research-sensemaking question was taken from Zhihu after google translation to keep the original presentations on the Zhihu platform.}
	\label{FIG: description_strategy}
\end{figure}

\begin{itemize}
    \item \textbf{Supplementing additional resources}. 
        
    \begin{itemize}
        \item \textit{Paper sources (38.6\%)}: Add the publication source of the research to support original paper tracing. A proportion of 17.3\% contained direct links to the e-print, and 21.3\% provided the title of the paper in the original language to help target the publication.        
        \item \textit{Supplementary materials (63.0\%)}: Users frequently attached links to supplementary materials, mostly linking to second-hand explanatory articles in Chinese and sometimes providing other related work, to provide more relevant information that helped users have a basic knowledge of the research. Note that \textit{supplementary materials} had a higher proportion than \textit{paper sources}, indicating the more frequent use of second-hand articles translating and interpreting the paper in Chinese than the original paper. Therefore, media and some individual domain experts, as typical authors of such second-hand articles, played an important role as mediators in such cross-language knowledge communication.

        \item \textit{Visuals} (9.9\%): Add images or videos (adapted from the original paper, second-hand articles, or drafted by the asker) to facilitate users' comprehension of research. Images (8.0\%) were more commonly used than videos (2.2\%), and only one question in the sample adopted both.
        
    \end{itemize}
    
    \item \textbf{Explaining comprehensive details}.
    \begin{itemize}
        \item \textit{Detailed methods} (37.3\%): Introduce the major methods used in the study (example shown in Figure \ref{FIG: description_strategy}).
        \item \textit{Detailed findings} (69.4\%): Detail the key findings of the study (example shown in Figure \ref{FIG: description_strategy}).
        \item \textit{Detailed questions} (9.9\%): Specify research-related questions, typically more detailed than the question title. For example, an asker proposed a question titled ``\textit{How do you understand the research claiming that long-term consumption of salt substitutes can reduce the incidence of cardiovascular diseases}'', and specified detailed questions in descriptions: ``\textit{What are the health risks of a high-salt diet? What is the difference between salt substitutes and regular salt? What are the health effects of regular consumption of substitute salt containing potassium?}''
    \end{itemize}

\item \textbf{Providing clear and engaging presentations}.
    \begin{itemize}
        \item \textit{Lede} (17.1\%): Use several opening sentences to introduce the research and entice users to continue reading, rather than directly describe research detail, e.g., ``\textit{\ul{How old is the Milky Way? How did it form and evolve?} The latest research published in Nature pointed out that the Milky Way may have gone through different stages of evolution...}''
        \item \textit{Structuring} (16.4\%): Use quotation or indentation to show a clear description structure that eases reading and helps users get the key points, as shown in Figure \ref{FIG: description_strategy}.
    \end{itemize}
\end{itemize}

\subsection{RQ2: Strategies' Correlations with Public Participation and Knowledge Construction}\label{RQ2-Findings}

The findings of RQ1 presented a comprehensive taxonomy of users' strategies in asking and describing research-sensemaking questions in CQA platforms. In this section, we demonstrate how these strategies correlated with user participation and knowledge co-construction in answers. The regression analysis on \textit{user participation} dimensions shed light on which strategies may attract more users' engagement (\textit{views, upvotes, followers} and \textit{answers}), and the regression analysis on \textit{knowledge co-construction} dimensions elucidated which question-asking strategies may lead to high-quality answers for knowledge construction - the answers that were on-topic (\textit{on-task discourse}), well-justified (\textit{evidence} and \textit{reasoning}), and based on socialized knowledge exchange and discussion (\textit{social}).

\noindent \textbf{Interpreting Regression Analysis}: This section presents regression models that reveal the correlation of question-asking strategies with user participation and knowledge co-construction in Table \ref{tab: reg_participation} and Table \ref{tab: reg_knowledge}. To ease the interpretation of effect size, we converted the regression coefficients to the ratio change of the dependent variable according to the specific link function of regression models. Particularly, for user participation dimensions, we report IRR (Incidence Rate Ratio) that denotes the rate ratio change of the dependent variable when increasing an independent variable by one unit ($\frac{y_{x+}}{y}$); for knowledge co-construction dimensions, we present OR (Odds Ratio) that indicates the odds change of the dependent variable when increasing an independent variable by one unit ($\frac{y_{x+}/(1-y_{x+})}{y/(1-y)}$). Therefore, both IRR and OR indicate positive correlations when their values exceed 1, with stronger positive correlations observed with larger values; conversely, IRR and OR below 1 indicate negative correlations, with stronger negative correlations observed with smaller values. 

\subsubsection{Descriptive Statistics}\label{RQ2-des}

\begin{figure}
	\centering
		\includegraphics[scale=.38]{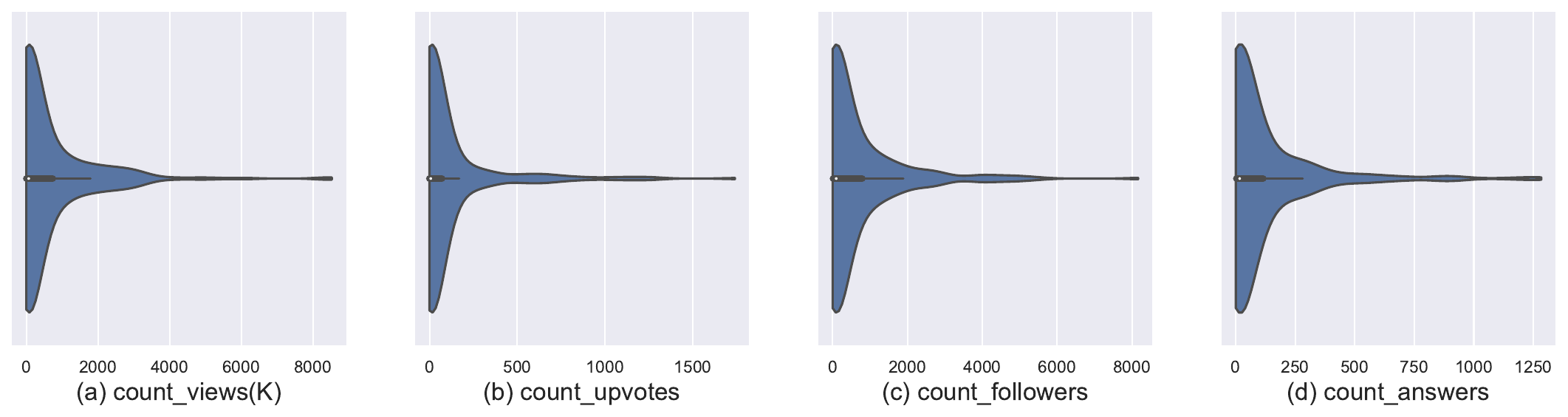}
	\caption{The violin plots that illustrate the long-tailed distribution of user participation in research-sensemaking questions.}
	\label{FIG: violin}
\end{figure}

Table \ref{tab: descriptive} presents the mean and standard deviation of dependent variables covering user participation and knowledge construction dimensions. Generally, research-sensemaking questions in the data sample attracted wide user engagement with 619,822 views and 109 answers on average. User participation dimensions also exhibited long-tailed distributions, which is demonstrated in Figure \ref{FIG: violin}. Based on the text classification on fine-grained knowledge construction dimensions, we found that research-sensemaking questions typically had a high proportion of answers with on-task discourse for research-related discussion (86.4\% on average). Nonetheless, only about half of the answers provided evidence such as data or theory to support their arguments, and only about 65\% answers were well-justified with logical reasoning. Socialized knowledge exchange and discussion, such as supplementing or rebutting, was less commonly observed (7.3\%) but was not negligible.

\begin{table}[htbp]
	\footnotesize
	\centering
	\caption{Mean and standard deviation of user participation and knowledge construction dimensions for research-sensemaking questions}
	\label{tab: descriptive}
	\begin{tabular}{p{1.3cm}|p{1cm}p{1cm}p{1cm}p{1cm}|p{1.2cm}p{1.2cm}p{1.2cm}p{1.2cm}}
		\hline
		& \multicolumn{4}{c}{User Participation (Count)}&\multicolumn{4}{c}{Knowledge Construction (Proportion)}\\
		\cline{2-9}
		& Views & Upvotes & Followers & Answers & On-task Discourse & Evidence & Reasoning & Social\\
		\hline

  		Mean & 619822.4 & 100.1 & 634.8 & 108.9 & 86.4\% & 50.7\% & 64.9\% & 7.3\%\\

  		Std Dev & 1151473.1 & 234.8 & 1115.4 & 201.6 & 0.18 & 0.28 & 0.25 & 0.13\\

	\hline
	\end{tabular}
\end{table}

\begin{figure}
	\centering
		\includegraphics[scale=.42]{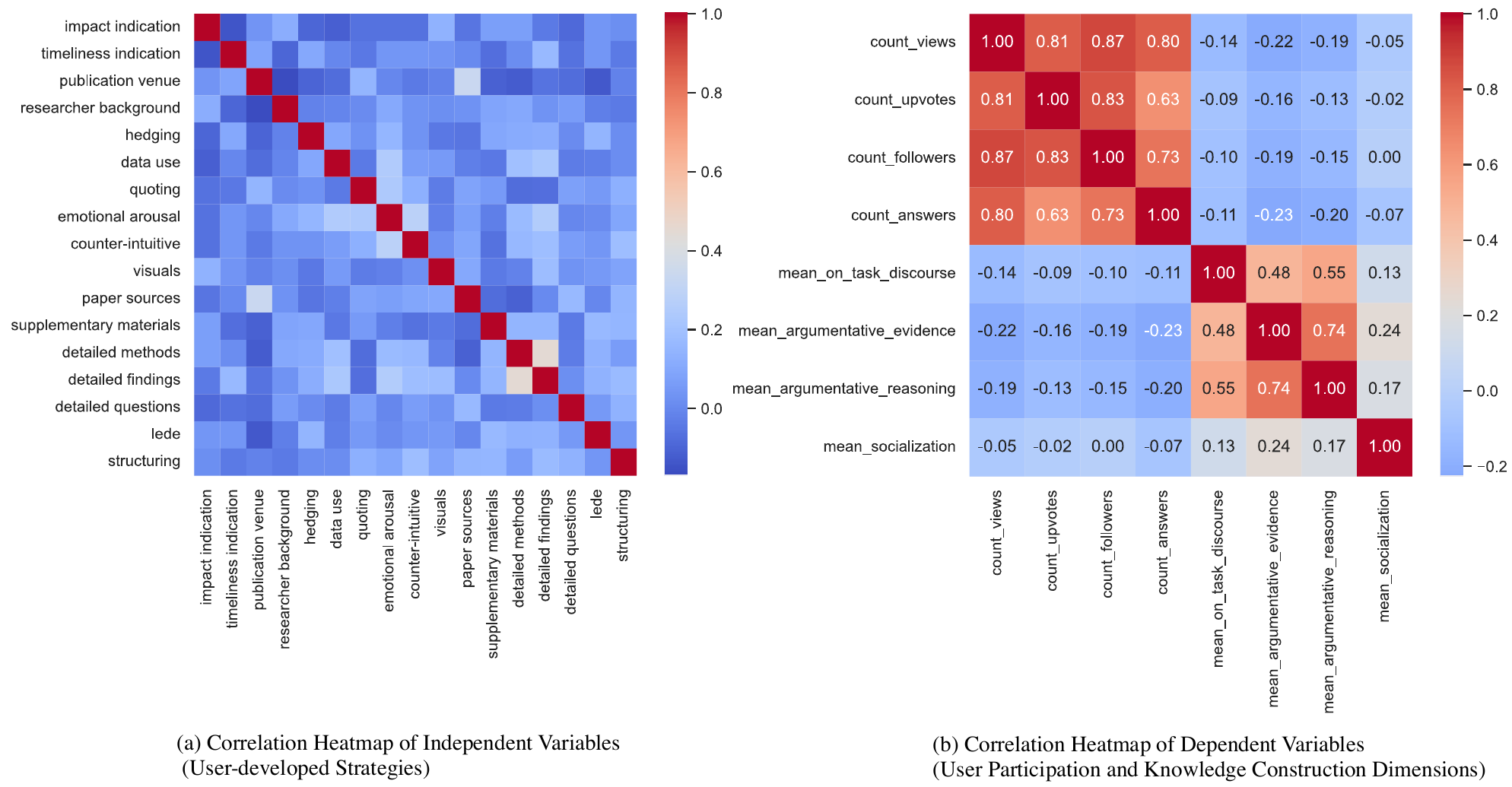}
	\caption{The correlation heatmaps of (a) user-developed strategies (the independent variables of regression analysis). The low correlation scores suggested no exclusions of strategies in regression analysis; and (b) user participation and knowledge construction dimensions (the dependent variables of regression analysis). It demonstrated negative correlations between user participation indexes and knowledge construction dimensions.}
	\label{FIG: correlation}
\end{figure}

Figure \ref{FIG: correlation} demonstrates the correlation heatmaps of user-developed strategies (independent variables) and user participation and knowledge construction dimensions (dependent variables). No pair of strategies in proposing research-sensemaking questions had a correlation score greater than 0.5 as described in Figure \ref{FIG: correlation} (a). Therefore, we did not exclude any strategies for regression analysis. Figure \ref{FIG: correlation} (b) shows that two variables among user participation indexes or knowledge construction dimensions generally had a  positive correlation, which was expected (e.g., questions with more views typically had more answers). However, it is important to note that \textbf{negative correlations were detected between user participation indexes and knowledge construction dimensions}. For instance, the correlation score between the number of answers and the proportion of evidence-supported answers was -0.23, which suggests that questions attracting more users to answer typically had a lower proportion of argumentative answers. This finding indicates the divergence between wide participation and epistemic and well-justified answer construction.

\subsubsection{Questioning Strategies' Associations with User Participation}\label{regResult-participation}

Table \ref{tab: reg_participation} shows the results of Poisson regression models predicting user participation in research-sensemaking questions across \textit{views (Model 1a), upvotes (Model 1b), followers (Model 1c)}, and \textit{answers (Model 1d)}. We highlight the following important findings:

\begin{itemize}

    \item Applying \textit{emotional} and \textit{counter-intuitive} narratives was correlated with wider user participation. Both the two narrative styles in question titles contributed to an increase of all user engagement indexes, especially for \textit{emotional arousal} feature which was associated with 52\% more views and 57\% more answers. It indicated that users might be attracted by questions with claims of research findings that triggered their emotional responses (e.g., anxiety or excitement) and were contrary to common-sense expectations (e.g., some controversial statements).

    \item Detailing \textit{researcher background} showed a positive correlation with wider user attention, resulting in over 20\% more views and over 10\% more upvotes, followers, and answers. Clarifying who conducted the research, such as the institutes of the research team and the leading scientists, might be more powerful to indicate the significance and establish the credibility compared to explicitly emphasizing the \textit{impact} or \textit{timeliness} of research, both of which did not show substantial promotion of user engagement.

    \item The factors gaining users' approval for the question quality (reflected from more ``upvotes'') may not align with the factors attracting more views and answers. For example, both question titles with \textit{data narrative} and \textit{quoting}, and question descriptions with \textit{visual representations} and \textit{supplementary materials}, received more upvotes from users, indicating users' appreciation of the question quality. Nonetheless, these features did not exhibit an elevation of views and answers.
    \item The efforts to present a \textit{clear and engaging question description}, including adding opening sentences as \textit{lede} and \textit{structuring} the description with indentation, positively correlated with user engagement in the research-sensemaking question with more upvotes, followers and answers.

    \item Demonstrating comprehensive details of research in question descriptions may not always work to augment user engagement. For instance, questions with \textit{detailed research methods} actually received 16\% fewer answers and \textit{detailed questions} also reduced answers by 9\%, potentially due to the higher linguistic barriers that hindered user participation.

    \item In addition to linguistic strategies in presenting the question titles and descriptions, \textit{research topics} also substantially correlated with user engagement (e.g., users' interest in social science-related research). Besides, questions asked by \textit{non-anonymous} accounts with \textit{more followers} received wider participation. Higher user engagement also manifested in questions with longer \textit{question titles} and more attached \textit{topics}, both of which were correlated with a 25\% increase in answer numbers.
\end{itemize}

\begin{table}[htbp]
	\footnotesize
	\centering
	\caption{Poisson regression models predicting user participation in research-sensemaking questions. IRR (Incidence Rate Ratio) indicates the ratio change of the dependent variable when increasing an independent variable by one unit. We color-coded features having significantly positive correlations with all dependent variables in \colorbox{green!25}{green}, and those having significantly negative correlations with all dependent variables in \colorbox{pink!25}{red}. ***p\textless0.001; **p\textless0.01; *p\textless0.05.}
	\label{tab: reg_participation}
	\begin{tabular}{p{2.4cm}p{2.4cm}p{0.5cm}p{1cm}p{0.5cm}p{1cm}p{0.5cm}p{1cm}p{0.5cm}p{1cm}}
		\hline
		& & \multicolumn{2}{l}{M1a: Views}&\multicolumn{2}{l}{M1b: Upvotes} & \multicolumn{2}{l}{M1c: Followers}&\multicolumn{2}{l}{M1d: Answers}\\
		\cline{3-10}
		& &IRR&Std. Err.&IRR&Std. Err.&IRR&Std. Err.&IRR&Std. Err.\\
		\hline
		
		\multicolumn{10}{l}{\textbf{Strategies in asking research-sensemaking question (question title)}}\\
		\hline
  
		\multirow{4}{1in}{Significance Signs} 
	    &impact indication&0.94***&0.002&0.95***&0.006&1.10***&0.002&1.01*&0.006\\
		&\crd timeliness indication&0.88***&0.003&0.92***&0.006&0.91***&0.003&0.84***&0.007\\
	    &publication venue&0.98***&0.003&\gr1.01&\gr0.007&1.05***&0.003&0.89***&0.007\\
            &\cgr researcher background&1.23***&0.002&1.10***&0.006&1.12***&0.002&1.13***&0.006\\
        \cline{1-2}
		\multirow{2}{1in}{Rigorous Descriptions} 
		&hedging&1.01***&0.002&1.02**&0.006&1.03***&0.002&\gr1.00&\gr0.006\\
            &data use&0.89***&0.003&1.02***&0.006&1.02***&0.003&0.97***&0.006\\
		
		\cline{1-2}
		\multirow{3}{1in}{Eyecatching Narratives} 
		&quoting&0.96***&0.003&1.08***&0.006&1.01***&0.003&0.88***&0.006\\
		&\cgr emotional arousal&1.52***&0.003&1.38***&0.007&1.25***&0.003&1.57***&0.007\\
  		&\cgr counter-intuitive&1.05***&0.002&1.08***&0.006&1.07***&0.002&1.08***&0.006\\

		\hline
		
		\multicolumn{10}{l}{\textbf{Strategies in describing research-sensemaking question (question description)}}\\

		\hline
    	\multirow{3}{1in}{Additional Resources}
		&visuals&-&-&1.06***&0.006&0.97***&0.002&0.92***&0.007\\
		&paper sources&-&-&\gr1.01&\gr0.007&1.10***&0.003&1.03***&0.007\\
		&supplementary materials&-&-&1.02***&0.006&0.94***&0.002&0.96***&0.006\\	
  
		\cline{1-2}
		\multirow{3}{1in}{Comprehensive details}
		&\crd detailed methods&-&-&0.84***&0.006&0.91***&0.003&0.84***&0.006\\
		&detailed results&-&-&1.07***&0.008&0.92***&0.003&1.02**&0.008\\
		&detailed questions&-&-&1.07***&0.005&1.06***&0.002&0.91***&0.006\\

            \cline{1-2}
		\multirow{2}{1in}{Clear presentations}
		&\cgr{lede}&-&-&1.13***&0.006&1.09***&0.002&1.10***&0.006\\
		&\cgr structuring&-&-&1.09***&0.006&1.16***&0.002&1.18***&0.006\\

		\hline
		\multicolumn{10}{l}{\textbf{Control Variables}}\\

            \hline
            \multirow{6}{1in}{Topic - science branch (ref: life science)} 
		&chemistry\&material&0.97***&0.003&0.87***&0.017&1.03***&0.003&0.84***&0.017\\
		&\cgr{earth\&space science}&1.06***&0.003&1.21***&0.008&1.16***&0.003&1.25***&0.008\\
		&\cgr{social science}&1.12***&0.003&1.11***&0.008&1.20***&0.003&1.41***&0.008\\
		&\crd computers\&technology&0.78***&0.010&0.90***&0.015&0.87***&0.006&0.80***&0.025\\
		&health&0.96***&0.004&1.18***&0.010&1.22***&0.004&1.12***&0.011\\
		&\cgr{math\&physics}&1.09***&0.002&1.36***&0.005&1.23***&0.002&1.18***&0.007\\

  		\cline{1-2}
		Topic - decision relevance &decision-related&0.96***&0.003&0.92***&0.007&0.91***&0.003&1.10***&0.006\\

		\cline{1-2}
		\multirow{3}{1in}{Goal (ref: only sensemaking)}
		&with assessing credibility&1.04***&0.002&1.04***&0.005&1.04***&0.002&0.96***&0.006\\
		&\crd with discussing implication&0.98***&0.003&0.88***&0.007&0.96***&0.003&0.88***&0.007\\
		&\crd with reasoning&0.98***&0.002&0.97***&0.006&0.96***&0.002&0.99*&0.005\\

		\cline{1-2}
		\multirow{3}{1in}{Asker Info}
		&\cgr{non-anonymous}&1.30***&0.003&1.49***&0.009&1.32***&0.003&1.10***&0.007\\
		&follower&1.09***&0.002&0.93***&0.006&1.07***&0.002&1.08***&0.005\\
		&\crd following&0.75***&0.004&0.61***&0.012&0.78***&0.003&0.71***&0.010\\

		\cline{1-2}
		\multirow{2}{1in}{Other Meta Info}
		&\cgr{question length}&1.20***&0.003&1.11***&0.007&1.22***&0.003&1.25***&0.007\\
		&\cgr{topic size}&1.31***&0.004&1.50***&0.011&1.32***&0.004&1.25***&0.008\\
		\hline
	\end{tabular}
\end{table}


\subsubsection{Questioning Strategies' Associations with High-quality Knowledge Co-construction}\label{regResult-knowledge}

Table \ref{tab: reg_knowledge} shows the results of Beta regression models predicting knowledge co-construction dimensions in research-sensemaking questions across (1) \textit{on-task discourse} (Model 2, epistemic dimension); (2) \textit{evidence} and \textit{reasoning} (Model 3a and 3b, argumentative dimension); and (3) \textit{social} (Model 4, social dimension). The regression analysis generated the following primary findings:

\begin{itemize}

    \item Rigorously describing scientific research may potentially attract higher proportions of quality answers. In particular, questions with \textit{hedging} were significantly correlated with reasoning ($OR=1.17$**) and social answers ($OR=1.15$*). Providing \textit{original paper links} of the research for sensemaking, as a valuable approach for users to find and interpret the original work, also received more high-quality answers in knowledge construction, which was positively associated with all epistemic ($OR=1.12$), argumentative ($OR=1.17$** for evidence and $OR=1.23$** for reasoning), and social dimensions ($OR=1.20$*). Besides, \textit{clear introduction (lede)} might also contribute to more on-task ($OR=1.20$*) and argumentative discourse ($OR=1.03$ for evidence and $OR=1.18$** for reasoning).

    \item Surprisingly, adding \textit{supplementary materials} in question descriptions correlated with more off-topic discussions ($OR=0.89$) as well as non-argumentative answers ($OR=0.91$* for evidence and $OR=0.89$* for reasoning). Through manual coding of a set of supplementary materials, we noticed that some supplementary materials appeared to be unprofessional news and articles created by media with extensive use of sensational expressions, which might suffer from exaggeration and misrepresentation and thus distract public attention.

    \item The \textit{researcher background} also negatively correlated with argumentative answers ($OR=0.89$* for evidence and $OR=0.88$* for reasoning). Note that this feature largely promoted public participation as revealed in Section \ref{regResult-participation}. We observed that as many questions tended to create contrast between famous researchers (or institutes) and controversial results for public attention, some answers were distracted to criticize the researcher rather than focus on the research-sensemaking task itself.

    \item More \textit{following} users of the asker positively predicted on-task ($OR=1.32$***) and argumentative answers ($OR=1.13$* for evidence and $OR=1.26$** for reasoning), which was potentially due to the askers' behavior of inviting relevant and high-quality contributors. Questions asked for \textit{discussing implication} had lower proportions of on-task answers ($OR=0.75$***) compared to other question-asking goals such as \textit{only sensemaking} (as the reference group) and \textit{credibility assessment} ($OR=1.14$*). Different \textit{science branches of topics} significantly associated with the social dimension in answers, indicating the varied collaboration levels of knowledge co-construction in different-discipline research-sensemaking questions.

    \item Different from their positive correlations with user participation, both \textit{question length} and \textit{topic size} negatively correlated knowledge co-construction dimensions. In particular, longer questions received lower proportions of on-task discourse ($OR=0.83$*). Also, even though adding more topics might attract broader users, \textit{topic size} had significantly negative correlations with the percentage of evidence-supported ($OR=0.90$*) or collaboratively-constructed ($OR=0.89$*) answers, possibly due to the contributions of non-expert users outside of the subject-related community.

\end{itemize}

\begin{table}[htbp]
	\footnotesize
	\centering
	\caption{Beta regression models predicting knowledge co-construction dimensions in research-sensemaking questions, including (1) on-task discourse (Model 2, \textit{epistemic dimension}); (2) evidence and reasoning (Model 3a and 3b, \textit{argumentative dimension}); and (3)social (Model 4, \textit{social mode dimension}). OR (Odds Ratio) indicates the odds change of the dependent variable when increasing an independent variable by one unit. We color-coded features having significantly positive correlations with at least two dependent variables in \colorbox{green!25}{green}, and those having significantly negative correlations with at least two dependent variables in \colorbox{pink!25}{red}. ***p\textless0.001; **p\textless0.01; *p\textless0.05.}
	\label{tab: reg_knowledge}
	\begin{tabular}{p{2.4cm}p{2.4cm}p{0.5cm}p{1cm}p{0.5cm}p{1cm}p{0.5cm}p{1cm}p{0.5cm}p{1cm}}
		\hline
		& & \multicolumn{2}{l}{M2: On-task}&\multicolumn{2}{l}{M3a: Evidence} & \multicolumn{2}{l}{M3b: Reasoning}&\multicolumn{2}{l}{M4: Social}\\
		\cline{3-10}
		& &OR&Std. Err.&OR&Std. Err.&OR&Std. Err.&OR&Std. Err.\\
		\hline
		
		\multicolumn{10}{l}{\textbf{Strategies in asking research-sensemaking question (question title)}}\\
		\hline
  
		\multirow{4}{1in}{Significance Signs} 
	    &impact indication&1.32***&0.086&\gr1.02&\gr0.052&\gr1.02&\gr0.065&\gr1.03&\gr0.076\\
		&timeliness indication&0.85*&0.073&\gr1.00&\gr0.050&\gr1.02&\gr0.066&\gr0.98&\gr0.072\\
	    &publication venue&\gr1.11&\gr0.075&\gr0.96&\gr0.053&\gr1.02&\gr0.065&\gr1.02&\gr0.074\\
            &\crd{researcher background}&\gr0.96&\gr0.075&0.89*&0.053&0.88*&0.064&\gr1.04&\gr0.070\\
        \cline{1-2}
		\multirow{2}{1in}{Rigorous Descriptions} 
		&\cgr{hedging}&\gr1.13&\gr0.075&\gr1.05&\gr0.049&1.17**&0.064&1.15*&0.071\\
            &data use&\gr1.00&\gr0.077&\gr0.93&\gr0.052&\gr0.90&\gr0.066&\gr1.10&\gr0.072\\
		
		\cline{1-2}
		\multirow{3}{1in}{Eyecatching Narratives} 
		&quoting&0.80**&0.077&\gr1.07&\gr0.051&\gr1.01&\gr0.065&\gr1.09&\gr0.075\\
		&emotional arousal&\gr0.96&\gr0.082&\gr0.91&\gr0.055&\gr0.97&\gr0.070&\gr1.01&\gr0.081\\
  		&counter-intuitive&\gr1.05&\gr0.073&\gr1.08&\gr0.050&1.17**&0.063&\gr1.00&\gr0.075\\

		\hline
		
		\multicolumn{10}{l}{\textbf{Strategies in describing research-sensemaking question (question description)}}\\

		\hline
    	\multirow{3}{1in}{Additional Resources}
		&visuals&\gr1.00&\gr0.072&\gr0.97&\gr0.049&\gr0.99&\gr0.063&\gr0.98&\gr0.071\\
		&\cgr{paper sources}&\gr1.12&\gr0.082&1.17**&0.055&1.23**&0.070&1.20*&0.077\\
		&\crd supplementary materials&\gr0.89&\gr0.079&0.91*&0.050&0.89*&0.065&\gr0.93&\gr0.072\\
  
		\cline{1-2}
		\multirow{3}{1in}{Comprehensive details}
		&detailed methods&\gr0.92&\gr0.078&\gr1.13&\gr0.054&\gr0.94&\gr0.069&\gr1.06&\gr0.077\\
		&detailed results&\gr0.92&\gr0.073&\gr1.01&\gr0.058&\gr0.99&\gr0.075&\gr1.02&\gr0.085\\
		&detailed questions&\gr1.12&\gr0.075&\gr1.08&\gr0.049&1.13*&0.064&\gr1.07&\gr0.068\\

            \cline{1-2}
		\multirow{2}{1in}{Clear presentations}
		&\cgr{lede}&1.20*&0.078&\gr1.03&\gr0.051&1.18*&0.067&\gr0.93&\gr0.074\\
		&structuring&\gr0.94&\gr0.079&0.90*&0.052&\gr0.89&\gr0.069&\gr0.97&\gr0.079\\

		\hline
		\multicolumn{10}{l}{\textbf{Control Variables}}\\

            \hline
            \multirow{6}{1in}{Topic - science branch (ref: life science)} 
		&\cgr chemistry\&material&\gr0.92&\gr0.066&1.12*&0.051&\gr1.09&\gr0.063&1.28***&0.054\\
		&\cgr{earth\&space science}&\gr0.94&\gr0.088&\gr0.96&\gr0.058&1.29***&0.078&1.20*&0.084\\
		&social science&\gr0.98&\gr0.100&\gr0.94&\gr0.066&\gr0.93&\gr0.085&1.37**&0.100\\
		&computers\&technology&\gr1.03&\gr0.080&\gr1.07&\gr0.054&\gr1.00&\gr0.070&1.23***&0.058\\
		&health&\gr1.03&\gr0.123&\gr1.02&\gr0.080&\gr1.14&\gr0.102&1.45**&0.121\\
		&math\&physics&\gr0.90&\gr0.081&\gr0.93&\gr0.054&\gr0.93&\gr0.070&1.21*&0.074\\

  		\cline{1-2}
		Topic - decision relevance &decision-related&1.28**&0.087&\gr1.09&\gr0.059&\gr1.06&\gr0.076&\gr0.89&\gr0.086\\

		\cline{1-2}
		\multirow{3}{1in}{Goal (ref: only sensemaking)}
		&with assessing credibility&1.14*&0.076&\gr1.00&\gr0.051&\gr1.04&\gr0.067&\gr1.03&\gr0.070\\
		&with discussing implication&0.75***&0.087&\gr0.99&\gr0.057&\gr0.91&\gr0.075&\gr0.98&\gr0.084\\
		&with reasoning&\gr0.97&\gr0.072&\gr0.98&\gr0.050&\gr1.03&\gr0.064&0.86*&0.075\\

		\cline{1-2}
		\multirow{3}{1in}{Asker Info}
		&non-anonymous&\gr1.03&\gr0.077&\gr1.02&\gr0.051&\gr1.09&\gr0.065&\gr0.96&\gr0.072\\
		&follower&\gr0.96&\gr0.069&\gr1.00&\gr0.051&\gr1.00&\gr0.063&\gr1.08&\gr0.071\\
		&\cgr following&1.32***&0.083&1.13*&0.053&1.26***&0.071&\gr1.00&\gr0.076\\

		\cline{1-2}
		\multirow{2}{1in}{Other Meta Info}
		&question length&0.83**&0.077&\gr0.97&\gr0.052&\gr0.91&\gr0.067&\gr0.92&\gr0.073\\
		&\crd{topic size}&\gr0.97&\gr0.077&0.90*&0.050&\gr\gr0.99&\gr0.065&0.89*&0.068\\
		\hline
	\end{tabular}
\end{table}

\subsection{RQ3: Users' Collaborative Work in Co-constructing Research-sensemaking Questions}\label{RQ3-findings}

RQ1 and RQ2 revealed the strategies in proposing research-sensemaking questions and their correlations with user engagement and knowledge construction. In this section, we took a further step to unpack the \textit{construction process} of research-sensemaking questions through collaborative editing. We illustrated the dynamics of users' collaborative work in co-constructing high-quality research-sensemaking questions, including \textit{co-establishing the topic scope}, \textit{scientific reframing and correction to enhance rigor}, \textit{co-contributing reliable and understandable external references}, and \textit{refinement for clear and engaging narratives}.

\subsubsection{Co-establishing the Topic Scope}

Adding or deleting topics, attached as tags of questions, has been identified as the most common co-editing behavior on the Zhihu platform~\cite{chen2020changing}. Through the log analysis, we found that such topic co-editing was a crucial process for users to co-establish the topic scope for research-sensemaking questions. It was a common practice of topic co-editing to exclude less relevant or redundant topics to refine the topic scope. For instance, in the question titled ``\textit{A survey shows that only 35\% of Chinese people sleep for more than 8 hours a day. Experts recommend going to sleep between 10pm and 11pm. How is your sleep quality?}'', the original asker used tags of ``\textit{health}'' and ``\textit{medical treatment}''. In the co-editing process, one user removed ``\textit{medical treatment}'', and another user supplemented ``\textit{sleep cycle}'' to better describe the topic. More importantly, users managed to balance the scientific topic scope for these questions based on topic co-editing, keeping them from being over-general without a focus, or getting limited into over-narrow and professional spaces that could not engage the public. For example, in the question titled ``\textit{Research has proved that a large number of white-tailed deer in the eastern United States have been infected with COVID-19. Will the COVID-19 infection of wild animals have an impact on the prevention and control of the pandemic?}'', the original asker attached topics of ``\textit{research}'' and ``\textit{health}''. Another user removed them and added a topic of ``\textit{SARS-CoV-2}'' to narrow down the topic to a more specific scope.

\subsubsection{Scientific Reframing and Correction to Enhance Rigor}

We observed that many original questions intentionally or unintentionally mistranslated, exaggerated, or misrepresented research findings such as omitting essential conditions. In this scenario, users' collaborative editing afforded scientific framing and correction to enhance rigor and rectify misinformation for some questions, which helped prevent public sensemaking in a misguided direction. For instance, an asker posted an original research-sensemaking question titled ``\textit{The Lancet published the latest inactivated vaccine data in Chile: Inactivated booster injections are 20-30\% less effective than mRNA vaccines. How to interpret it?}''. Another user supplemented an important condition of the research, ``\textit{with the basis of two doses of inactivated vaccine}'', to make the statement more accurate and rigorous. Such scientific reframing was not limited to the quoted research findings, but also applied to the meta information about the research. For example, when the original asker posted ``\textit{What do you think of Nature's \ul{comment}: In praise of replication studies and null results?}'', another user changed the ``\textit{comment}'' to ``\textit{editorial}'' to correctly describe the article type. In some other examples, scientific reframing was also reflected in ways such as adding the specific time the research was conducted and published for time-sensitive work, highlighting the studied populations, and changing approximated numbers to specific numbers.

\subsubsection{Co-contributing Reliable and Understandable External References}

The original paper and its supplementary interpretations intuitively played a significant role in research sensemaking, which was also validated in RQ1 and RQ2. The log analysis further revealed the prevalence of collaborative contribution on the external references for the question, including links to both the original paper and supplementary materials. Some research-sensemaking questions initially lacked links to the original paper, and other domain experts added them based on the research description. It was surprising to note that even when the original asker had provided external links to the original paper or relevant materials, other users might supplement references that were more understandable or reliable. For instance, in a question where the original asker attached the link to the paper in English, another user supplemented a Chinese article from domain experts translating and interpreting the paper in a more comprehensible way, aiming to engage a wider audience. In another example, the original question provided a post on social media Weibo discussing the research, and another user later modified it to a more authoritative Chinese article summarizing relevant research findings. These examples indicated users' collaborative efforts to establish external references that could elucidate the research to a broad audience.

\subsubsection{Refinement for Clear and Engaging Narratives}

We found that some narrative strategies aiming for clear and engaging presentations appeared to be the results of collaborative refinement. For example, the strategy of \textit{quoting key findings} in question titles was frequently implemented in collaborative editing by voluntary contributors, who added quotation marks for the key findings to highlight them. Some users also volunteered to add one or several opening sentences as the \textit{lede} for the question descriptions that initially began with direct research explanations, some just simply as ``\textit{a recent study published in <publication venue> has attracted widespread attention. It is a collaborative research project conducted by <institutes>}''. These efforts of collaborative refinement manifested unwritten norms that were commonly accepted by users as ``good practice'' for asking research-sensemaking questions.

\section{DISCUSSION}

This work identifies strategies used in composing research-sensemaking questions that are associated with engaging with the audience and eliciting quality answers from the community. Generally, this work unveils a new pattern of participatory science communication with great potential to facilitate knowledge dissemination and sensemaking, as well as points out its existing challenges especially the gap between wide public participation and quality answer construction. This section situates the findings within the literature, reflects on the opportunities and challenges for CQA-based research-sensemaking, and proposes design implications to facilitate accurate, engaging, and effective science communication.

\subsection{From Passive Receivers to Active Askers: Rethinking Users' Efforts in Initiating Research-sensemaking}

The participatory knowledge exchange, with a focus on dialogue and public engagement, has characterized contemporary science communication on social media~\cite{williams2022hci}. It goes beyond the traditional model of linear transmission from scientists to media and then public~\cite{bruggemann2020post,chilvers2010sustainable}, which necessitates rethinking the participatory practice of stakeholders. This study deepened the understanding of an emerging role of the public in science communication, i.e., the question-askers who initiated research-sensemaking on CQA platforms. In this section, we discussed how users were engaged in this process and how it afforded new opportunities for science communication.

Effectively engaging the general public with science-related information has been a challenging task in science communication even for mature science communicators like scientists~\cite{gero2021makes}. This work uncovered a comprehensive taxonomy of users' strategies in proposing research-sensemaking questions in CQA platforms, shedding light on \textbf{public wisdom in initiating research-sensemaking} when their roles became askers. Most user-developed strategies naturally reflected practical principles established in prior work. For example, the SUCCESS framework proposed by Finkler and Le{\'o}n emphasized the significance of being Simple, Unexpected, Concrete, Credible, Emotional, and Science Storytelling for science-related rhetoric~\cite{finkler2019power}. Resonating with it, users crafted strategic research-sensemaking questions that noted researcher background and publication venue to establish \textit{credibility}, used data narratives to enhance \textit{concreteness}, and applied counter-intuitive and emotional statements for \textit{unexpectedness} and \textit{emotionalization}, some of them managing to attract millions of views. As shown in Section \ref{RQ3-findings}, some strategies even developed as unwritten norms collaboratively implemented by voluntary community members. These findings contributed new empirical evidence to strategic science communication in real-world settings~\cite{flemming2018emotionalization, finkler2019power,dahlstrom2014using}. More importantly, they added new nuances on how users naturally integrated the art of questioning and narratives for knowledge exchange in participatory science communication~\cite{liang2019scientists}. On this note, we suggest future work broadly investigate how to empower the general public as active initiators in knowledge exchange, and exploit the opportunities of ``speaking the language of the general public'' for science communication.

Nonetheless, limitations of users' spontaneous questioning for research sensemaking also emerged. Some more professional science communication strategies for rigor and accuracy, such as the use of visual representations to describe the research, were less observed in our dataset. Meanwhile, we noticed that part of user-developed strategies such as counter-intuitive and emotional expressions potentially introduced distortions, exaggerations or misrepresentations of the research outputs, leading to misguided research sensemaking. It could contribute to polarized discourse, a feature of ``post-normal'' science communication on digital media~\cite{bruggemann2020post}. Such misinformation in research-sensemaking questions might also be unintentionally brought by mistranslation and missed essential conditions when non-experts lacked specific domain knowledge. Therefore, we suggest the involvement of scientists and experts to assist with question crafting to improve its quality and upskill question askers, e.g., affording a mentorship program for scientific questioning~\cite{ford2018we}. The deployment of flagging mechanisms~\cite{li2022flagging} for research-sensemaking questions, leveraging AI-supported or crowdsourced credibility checking to mark potentially unreliable or misguided questions, is also a promising direction.

\subsection{Questioning Strategies Failed to Ensure Both: Understanding the Tension Between Wide Participation and Quality Knowledge Construction}

The existing literature in science communication on social media has indicated the dilemma of engaging the general public and ensuring the rigor of knowledge exchange~\cite{august2020explain,gero2021makes,williams2022hci}. Epistemic science discussion intrinsically correlates with linguistic barriers (e.g., scientific terminology and hedge words) that gatekeep public participation~\cite{august2020explain}, while public science discussion without the constraints of rigor and objectivity may exploit research outputs for conspiracy theories and extremist ideology~\cite{yudhoatmojo2021we,lee2021viral}. By analyzing different knowledge co-construction dimensions on research-sensemaking questions, this work unearthed more nuanced dynamics of \textit{the tension between wide participation and quality knowledge construction} in science communication, and proposed design implications to cope.

\subsubsection{The Dilemma between Wide Participation and Quality Knowledge Construction}

The content analysis of the answers to research-sensemaking questions provided empirical evidence of the divergence between broad public participation and high-quality science discourse. As revealed in Section \ref{RQ2-des}, all public participation indexes of questions (views, votes, followers and answers) had varied degrees of negative correlations with the proportions of high-quality knowledge construction dimensions (on-task discourse, argumentative claims with evidence and reasoning, and socialized knowledge discussion). This finding validated the difficulty of balancing the quantity of public participation and quality of knowledge construction~\cite{august2020explain,zhang2023understanding}. In fact, such tension reflects a potential challenge under the blurring boundaries of science and journalism in the era of social media~\cite{bruggemann2020post}. Communicating science to the public no longer necessarily relies on professional intermediaries as gatekeepers; instead, the participation of ``scientist citizens'' and contextual interpretation of science characterize such post-normal science communication~\cite{pietrucci2019scientist}. On this note, being ``newsworthy'' may become a more salient feature for public research sensemaking rather than being ``scientific'', along with which misguidance and misinterpretation may prevail~\cite{yudhoatmojo2021we,lee2021viral}. Therefore, more investigations are warranted to afford more specialized guidance interfaces for the public to contribute high-quality knowledge, and develop more thorough moderation mechanisms to mitigate off-topic, polarized, and misleading discourse in addition to affording accessibility~\cite{burns2003science}.

This work further demonstrated how question-asking strategies that attracted more public participation may not align with strategies facilitating epistemic and argumentative knowledge establishment. For example, Section \ref{RQ2-Findings} revealed that the strategy of highlighting \textit{researcher background} correlated with more views and answers, but diverted the focus of some answers away from sensemaking to commenting institutes or researchers. In contrast, \textit{detailed questions}, typically calling for more specific and professional answers, correlated with high-quality knowledge construction but not more answers. These findings emphasized the significance of curating research-sensemaking questions that considered its influence on both engaging a wider audience and stimulating responses with higher quality. How questions were proposed largely influenced the public engagement and sensemaking directions in CQA platforms, which was rather different than tweetorials that disseminated knowledge themselves~\cite{gero2021makes} or more targeted questions with specific receivers in a community~\cite{gruzd2020coding}. Also, the discrepancy between wide participation and rigorous sensemaking necessitates nuanced investigations focusing on fine-grained dimensions measuring public engagement on science-related topics rather than only numerical indexes, which echoes prior work~\cite{xia2022millions,williams2022hci}.

\subsubsection{Design Implications}

Notably, we highlight the crucial role of the sociotechnical context in shaping the trade-off. A widely applicable and long-lasting value in knowledge co-creation, rather than one-off and personal questions, has gradually become the focus of CQA platforms~\cite{liu2023coargue,zhang2019understanding,anderson2012discovering}. Based on it, the platform design such as voting mechanisms and recommendation algorithms may encourage widely applicable questions and answers acknowledged by more users rather than very specific ones~\cite{anderson2012discovering,asaduzzaman2013answering} (also consider the removal of question details in Quora~\cite{QuoraDescription}). This incentive also intersects with the monetization model of CQA platforms~\cite{fang2021post}, which usually motivates askers/answerers to prioritize wide attention for potential profits. However, it may inadvertently conflict with certain research-sensemaking questions that require specificity to achieve rigor. For example, askers may omit specific descriptions or conditions that are actually crucial for particular research to attract wider attention. Moreover, with the aim of gaining wide engagement, users may intentionally pose misguided research-sensemaking questions, such as misinterpreting research with counter-intuitive conclusions, just setting a target for public emotional venting; or selectively reporting controversial findings only to initiate off-topic debates. Therefore, we suggest a knowledge-related feedback mechanism, in addition to upvotes, as an alternative feedback and evaluation system for research-sensemaking Q\&A. For example, verified domain experts or established users on related topics may endorse questions or answers with knowledge-related tags (e.g., ``experimental evidence'' and ``reasoning with data''), and other users could show their approval of them. The reputation and recommendation systems may take these dimensions into consideration to spur quality knowledge construction. 

Besides, when broad participation on science topics inevitably introduces off-task discourse, we suggest CQA platforms afford options to ease the seeking for quality knowledge-related answers, so as to facilitate those aiming to figure out the research (in contrast to users with an entertaining or conversational goal). The aforementioned knowledge-related tags provide one potential way to assist with quality answer filtering. Another possible approach is a specialized AI-supported sorting and filtering interface, through which users could sort answers by AI-evaluated relevance, argumentative claims, or socialized discussions. The good performance of automatic classifications on these knowledge construction dimensions in this work suggests the feasibility of this interface. Nonetheless, though the AI-supported sorting and filtering interface might ease research sensemaking and spur quality answers, further investigations are warranted to understand potential limitations such as algorithmic biases and possible risks of crowd gaming algorithm~\cite{epstein2020will}. 

Finally, it is a promising direction to utilize the power of social connections for constructive research sensemaking, such as enhancing the reward mechanism for high-quality answer inviting. Our findings also noted the influence of social networks on answer quality, when questions proposed by askers following more users (and thus potentially inviting more relevant and quality contributors) had more epistemic and argumentative discussions.

\subsection{Proposing Research-Sensemaking as a Collaborative Effort: Unpacking Opportunities and Challenges of Co-editing Research-Sensemaking Questions}

The affordance of collaborative question editing, similar to Quora~\cite{maity2018analyzing} and Stack Overflow~\cite{li2015good}, reflected the attempts of leveraging collective wisdom to construct high-quality questions on Zhihu~\cite{chen2020changing}. This section discusses how collaborative editing brings opportunities and challenges to the specific setting of research-sensemaking.

When presenting accurate and engaging science-related information naturally puts a high demand on questioning techniques, we note that collaborative editing plays an irreplaceable role in constructing and refining research-sensemaking questions. Section \ref{RQ3-findings} demonstrates the value of co-editing ranging from setting up a suitable topic scope, establishing reliable and comprehensible external information, to enhancing the rigor and clarity of presentations. These findings enrich the understanding of the potential of collaborative editing in scientific questioning~\cite{maity2018analyzing,li2015good}. When constructing high-quality research-sensemaking might be challenging for non-experts, the involvement of collective wisdom may not only be beneficial to generate more engaging and rigorous research-sensemaking questions, but also work as a descriptive influence of norms ~\cite{cialdini1990focus,august2020explain} that guide newcomers to ask in an acceptable way. It could also alleviate unique challenges of science communication, such as moderating misuse and polarization~\cite{yudhoatmojo2021we,lee2021viral} and establishing credibility~\cite{bucchi2017credibility} with collaborative work, especially when only ``credible'' users were qualified to co-edit questions on Zhihu~\cite{chen2020changing}. To this end, it is warranted to investigate how the power of collaborative editing could be adapted to other community-based or social network-based science communication settings.

On the other hand, co-editing research-sensemaking questions inherit and develop some issues of collaborative editing. Though the relatively high barrier of scientific language made ``editing war''~\cite{kittur2007he} less frequently observed in our sample, it also complicated the justification of editing~\cite{chen2020changing}. Many users used the default reasons such as ``punctuation and formatting error'' that did not point out the essential rigor-related problems and might not convince the original askers. On this note, we suggest some lightweight design improvements such as adopting specialized editing terminologies for science-related topics, or automatically examining edits to provide suggestions on justification reasons. Besides, some reframing on subjective narratives (e.g., detailed personal questions in the description) faces the tension between the subjective nature of question-asking and the objective nature of science communication. Though one reason for introducing collaborative editing on Zhihu is to generalize some questions~\cite{ZhihuCollaborativeEditing}, it may harm specific users' interest in research-sensemaking and limit more narrow directions of knowledge exchange. Therefore, we call for future investigations on question redirection and branching systems to make science communication more accessible to both general and personal research-sensemaking questions.

\subsection{Limitations}

This work has the following limitations:  (1) We used a keyword-based data collection approach, which could promote the purity and representativeness of the dataset but inevitably limited the dataset size and diversity (e.g., questions implicitly proposing research-sensemaking were largely excluded it our dataset); (2) We adopted regression analysis to capture how question-asking strategies were correlated with user-participation or knowledge construction dimensions. Even though the results were valuable in helping understand the potential influence of question-asking strategies, the correlations are not sufficient to claim causal relationships as a general limitation of regression analysis; (3) We measure the potential effects of questioning strategies only from a quantitative perspective (i.e., how they correlated with the \textit{observed} user participation and knowledge construction). A qualitative view focusing on how users perceived and responded to these questioning strategies (e.g., through interviews) was also significant to understanding this science communication setting; (4) Due to the scope of this study, we did not pay close attention to some more nuanced research questions that emerged from the analysis, such as a comprehensive understanding of how the tension between wide participation and quality knowledge construction appeared, how askers perceived and balanced this trade-off, and how misrepresented or exaggerated questions misguided the knowledge construction in answers. We suggest future work to systematically investigate these specific challenges of CQA-based science communication to facilitate accurate and effective knowledge construction and dissemination online. Besides, though this work contributes to the HCI and CSCW literature by expanding the understanding of science communication in a non-western context, a cross-platform or cross-cultural comparative study would be beneficial to unpack the sociocultural factors in science communication and provide valuable insights into CQA-based research sensemaking.

\section{CONCLUSION}

Users are increasingly gathering in research-sensemaking questions to discuss science-related topics in CQA platforms, and good question-asking strategies are crucial to attract public participation and facilitate knowledge construction. This work makes the first attempt to investigate strategies in crafting research-sensemaking questions, their correlations with user engagement and knowledge construction, and co-editing efforts from the community to implement them. To achieve this, we collected 837 science-related questions with 157,684 answers and conducted a mixed-methods study on the Zhihu platform. Through an open coding approach, we captured a comprehensive taxonomy of user-developed strategies in question titles and descriptions to enhance rigor (e.g., hedging and data use) and engage the audience (e.g., emotional and counter-intuitive statements). Through regression analysis, we identified these strategies' correlations with public participation and knowledge construction, such as the increased views and answers in emotional questions. It helped to unpack the divergence between wide participation and argumentative knowledge construction when few strategies could promote both. Finally, the inductive log analysis suggested the unique values of collaborative editing in promoting question quality, such as collaborative reframing and correction to enhance rigor. We discuss design implications for accessible, accurate, and effective research-sensemaking in CQA platforms.

\bibliographystyle{ACM-Reference-Format}
\bibliography{sample-base}

\end{document}